%% file: kri16.tex
\documentclass[openacc]{rstransa}
\usepackage[export]{adjustbox}
\usepackage[utf8]{inputenc}
\usepackage{amsmath}
\usepackage{amsfonts}
\usepackage{amssymb}
\usepackage{stackengine}
\usepackage{graphicx}
\usepackage{subfig}
\usepackage{xcolor}
\usepackage{float}

\usepackage{color} 
\newcounter{commentzaehler}

\jname{rsta}
\Journal{Phil. Trans. R. Soc.}

\begin{document}

\title{Synchronization Patterns: From Network Motifs to Hierarchical Networks}

\author{
Sanjukta Krishnagopal$^{1,2}$,
Judith Lehnert$^{1}$,
Winnie Poel$^{1}$,
Anna Zakharova$^{1}$ and
Eckehard Schöll$^{1}$}

\address{
$^{1}$
Institut für Theoretische Physik, Technische Universität Berlin,
10623 Berlin, Germany
\\
$^{2}$ Department of Physics, Birla Institute for Technology and Science-Pilani, Goa-403726, India

}
\subject{Non Linear Dynamics, Applied Mathematics}

\corres{Sanjukta Krishnagopal\\
\email{sanju33@gmail.com}}
\begin{abstract}
We investigate complex synchronization patterns such as cluster synchronization and
partial amplitude death in networks of coupled Stuart-Landau
oscillators with fractal connectivities. The study of fractal or self-similar topology is
motivated by the network of neurons in the brain. This fractal
property is well represented in hierarchical networks, for which we present three different
models. In addition, we introduce an analytical eigensolution method and provide a comprehensive picture of the
interplay of network topology and the corresponding network dynamics,
thus allowing us to predict the dynamics of arbitrarily large
hierarchical networks simply by analyzing small network motifs. We also show that oscillation death
can be induced in these networks, even if the coupling is symmetric,
contrary to previous understanding of oscillation death. Our results
show that there is a direct correlation between topology and dynamics: Hierarchical networks exhibit the corresponding
hierarchical dynamics. This helps bridging the gap between mesoscale motifs and macroscopic
networks.
\end{abstract}

\maketitle

\section{Introduction}

In the last decades, synchronization and its control has sparked tremendous scientific
interest in network science because of its wide applicability \cite{SCH16}. 
Examples of synchrony range from genetic oscillators \cite{GAR04}
and population dynamics \cite{BLA99a} via data mining \cite{MIY07} and
power grid networks \cite{WIT12,ROH12} to opinion formation
\cite{PLU05}. While early research focused on in-phase (or zero-lag)
synchronization, recently more complex synchronization patterns such as
group or cluster \cite{GOL02a,SOR07,KES08,KAN11,ILL11,DAH12,LUE12a,BLA13,WIL12a,WIL13,
PEC14,ROS15,SOR16a}, or partial
synchronization \cite{MOH06,WAG02a,RUB02,POE15} have moved towards the center of scientific
interest -- in theoretical studies as well as in experiments. Partial
synchronization describes a state where a part of the network is in
synchrony -- this can be in-phase, cluster, or group synchronization
-- while other parts exhibit oscillation quenching, i.e., amplitude death or oscillation death \cite{ATA02a, SAX12, KOS13}, or oscillate incoherently. 
The spatial coexistence of coherent and incoherent dynamics in a network of identical elements is called a chimera state
\cite{KUR02a,ABR04,HAG12,TIN12,MAR13,LAR13,OME13,ZAK14,PAN15,OME16,SEM16,SCH16b}.

Amplitude death is associated with the stabilization of an already
existing trivial steady state, while oscillation death is
characterized by an inhomogeneous steady state which is induced by the
coupling. Applications of  amplitude death pertain mainly to controlling physical and
chemical systems (e.g., coupled lasers) \cite{KUM08} and suppressing
neuronal oscillations \cite{ERM90, CAK14}, while oscillation death has been suggested
as a mechanism to generate heterogeneity in homogeneous systems, e.g.,
stem cell differentiation in morphogenesis \cite{SUZ11}. So far oscillation death has
been described as the result of symmetry breaking \cite{KOS13,ZAK13a,SCH15b}. Here,
we demonstrate that oscillation death can also occur in symmetric networks with symmetric
coupling when the collective frequency of the oscillators tends to
zero.

On the topological side, we focus on hierarchical networks exhibiting
a fractal or self-similiar structure \cite{PRO12,OME15,HIZ15,ULO16,TSI16}, which is
motivated by the intricate architecture found in neural networks:
Diffusion Tensor Magnetic Resonance Imaging (DT-MRI) results show that
neurons in the mammal brain are far from being linked homogeneously
but are connected in a
fractal manner, with fractal dimensions varying between 2.3 and 2.8,
depending on local properties, on the subject, and on the noise
reduction threshold\cite{KAT09,EXP11,KAT12,KAT12a,PRO12}. Other examples of
fractal networks  include  protein interaction networks\cite{SON05a} or biochemical
reactions in the metabolism\cite{RAV02} and hyperlinks in the World Wide Web\cite{DIL02}.

Ground-breaking work connecting the topological properties of a
network with its dynamics has been done by Pecora et al. in
introducing the master stability function for the completely
synchronous state \cite{PEC98}. This method has been extended to
time-delayed coupling \cite{DHA04,CHO09,FLU10b,SOR13,WIL14} and group
and cluster synchronization \cite{SOR07,TAY11,DAH12,PEC14}. However, the master
stability function considers the network topology on a global scale,
while very little is understood about how the global dynamics is
influenced by the network motifs, i.e., how the topology on a
mesoscale level influences the dynamics of the whole network. A first
step in this direction has been made by Do et al. establishing  that certain
mesoscale subgraphs are of crucial importance for the global dynamics
of the network\cite{DO12}. An analytical method for studying partial synchronization states in mesoscale motifs has been presented in \cite{POE15}. Here, we show how, in hierarchical networks, we can analytically predict the
global dynamics from the topology of the small motifs by extending the
eigensolution concept suggested in \cite{POE15}.

As a model, we consider coupled Stuart-Landau oscillators, a
paradigmatic normal form that naturally arises in an expansion of oscillator
systems close to a Hopf bifurcation. Networks of Stuart–Landau
oscillators have been used as oscillator models for neural networks
\cite{ TAN14}. Fairly simple topologies  of coupled
Stuart–Landau oscillators such as all-to-all networks \cite{MAT90,ERM90,HAK92,ATA03} and regular arrays of Stuart–Landau oscillators \cite{MIR90}
have been studied in several works. Here, we couple these oscillators in a more complex
hierarchical topology. To this end, we introduce and compare three
different models of hierarchical networks. We also explain, using the analytic
eigensolution concept, the coupling induced transition between cluster state dynamics and oscillation death.

The paper is organized as follows: In Sec.~\ref{model}, we
introduce the Stuart-Landau oscillator. In Sec.~\ref{analytic}, we
elaborate on the eigensolution method used to find analytical
solutions. We then present and briefly discuss
the network topology of three different models in Sec.
\ref{top}. Section~\ref{results} is organized into three subsections,
each of which presents the numerical and analytical results for the three
models, respectively. Finally, in Sec.~\ref{concl}, we provide a
conclusion and discuss future directions of research.

\section{Model}
\label{model}

In this paper, we study the Stuart-Landau oscillator, a generic model for a system close to a Hopf bifurcation. The dynamics of the $i^{\text{th}}$ oscillator, $i \in \{1, . . . ,N\}$, is given by
\begin{equation}
\dot{z_i} = f(z_i) + \sigma e^{\mathrm{i}\beta}\sum_{j=1}^NA_{ij}z_j,
\label{eq:f}
\end{equation}
\begin{align}
f(z)&=(\lambda+ \mathrm{i}  \omega-|z|^2)z,\label{eq:ff}
\end{align}
where $ z \in \mathbb{C}$ and $\lambda, \omega, \sigma, \beta \in \mathbb{R}$,
$\omega$ is the oscillator frequency. In the uncoupled oscillator
($\sigma=0$), $\lambda$ is the bifurcation parameter: For $\lambda > 0$, a
limit cycle of radius $\sqrt{\lambda}$ exists that is born in a supercritical
Hopf bifurcation at $\lambda = 0$. The
parameters $\sigma$ and $\beta$ are the coupling strength and coupling
phase, respectively.

The oscillators are connected via an instantaneous coupling as given by the
adjacency matrix $A_{ij}$. We normalize the adjacency matrix to 
unity row sum, i.e., $\sum_{j=1}^NA_{ij} = 1$, ensuring the existence
of an invariant synchronization manifold. The three models specifying
the topology of the network, and therefore  $A$, are introduced and discussed in detail in Sec.~\ref{top}. 

\section{Analytical eigensolution}
\label{analytic} 
\begin{figure} 
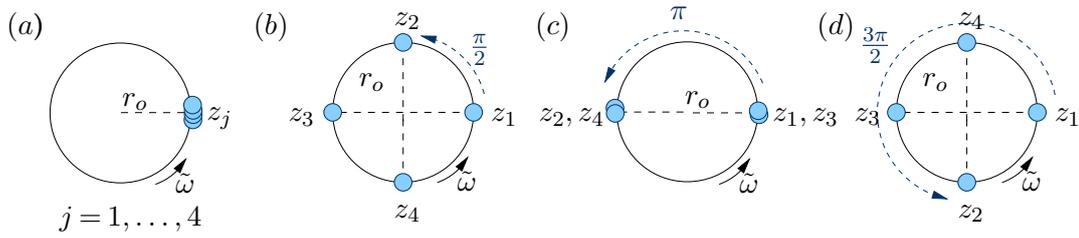
\caption{Schematic view of (a) zero-lag  synchronization ($j=0$, $M=1$), (b) the splay state  ($j=1$, $M=4$), (c) 
antiphase-synchronization  ($j=2$, $M=2$), and (d)  the reversed splay state
($j=3$, $M=4$). Each cluster consists of the same number of nodes.\label{fig:schema}} 
\end{figure}

In \cite{POE15} an analytical eigensolution approach
was suggested to determine the in-phase synchronized,
antiphase-synchronized, and amplitude death solutions of
Eq. \eqref{eq:f}. Here, we will give a brief summary of the results
obtained in \cite{POE15}. While doing so, we will also generalize this
eigensolution concept to cluster synchronization -- states where all nodes oscillate with the same amplitude $r_o$ and the same frequency $\tilde{\omega}$ but are
organized in equally sized clusters with a constant phase lag of
$\frac{2\pi j}{N}$ ($j \in \{1, \ldots, N\}$ where $N$ is the total number of nodes) between consecutive nodes\cite{CHO09}.  For $j>0$,
the number of clusters can be calculated as $M=lcm(j,N)/j$, where lcm
stands for the least common multiple. $j=N$ corresponds to in-phase
synchrony, while $j=N/2$ denotes the antiphase-synchronized state (for even $N$).  In
general, depending on $N$ and $j$, several states might exist which
are characterized by the same number of clusters.
Figure~\ref{fig:schema} shows a schematic view of all cluster states
in a unidirectional ring configuration of four nodes. In panel (a),
the nodes are in zero-lag synchronization. Panels (b) and (d) show two
different splay states, i.e., $M=N=4$, where the phase difference
between subsequent nodes in panel (b) is $\pi/2$, i.e., $j=1$, and in
panel (d) $3\pi/2$, i.e., $j=3$. Panel (c) depicts the anti-sychronized state.

The eigensolution concept is based on the idea that each eigenvector  $v$
of the adjacency matrix, with components $v_i$, fulfilling
\begin{equation}
|v_i|\in \{0,1\},\forall i=1, \ldots, N, \label{eq:1}
\end{equation}
corresponds to a solution, i.e., each component of $v$
 is either zero or a complex root of unity. This seems to
be a rather strong restriction on the eigenvector. However, the eigenvectors
of all hierarchical network topologies considered in this paper fulfil
this condition, and hence this method can directly be extended to all
our models (described in Sec.~\ref{top}).  Furthermore, all
eigenvectors of circulant adjacency matrices are known to have
eigenvectors with components equal to complex roots of
unity. Circulant matrices are of great current interest in the study of
chimera states\cite{OME11,OME12,ZAK14,OME16} because the corresponding
topology is invariant under discrete rotations (dihedral symmetry group).

We use the ansatz $z_i =v_i z_\eta$ in Eq.~\eqref{eq:f}, where $\eta$ denotes the
eigenvalue of the adjacency matrix corresponding to the 
eigenvector $v$.
Using Eq. \eqref{eq:1} , we can write $f(v_i z_\eta)=
v_i f(z_\eta)$. Thus, Eqs. \eqref{eq:f} decouples to
\begin{equation}
\dot{z_\eta}v_i = f(z_\eta) v_i + \sigma e^{\mathrm{i}\beta } \eta z_\eta v_i.
\end{equation}

Dividing by $v_i$ and introducing
\begin{equation}
\sqrt{\tilde{\lambda}} \equiv \sqrt{\lambda + \eta_r \sigma \cos \beta
  - \eta_c \sigma \sin \beta}, \quad \tilde{\omega} \equiv \omega +
\eta_r \sigma \sin \beta + \eta_c \sigma \cos \beta. \label{wtilde}
\end{equation}
where $\eta_r, \eta_c$ are the real and complex parts of $\eta$, respectively, we obtain
\begin{equation}
\dot{z_\eta}= (\tilde{\lambda}+\mathrm{i}\tilde{\omega}-|z_\eta|^2)z_\eta.
\end{equation}

This has the form of a decoupled Stuart-Landau oscillator $z_\eta$, the solution for which is known to be $z_\eta= \sqrt{\tilde{\lambda}} e^{\mathrm{i}\tilde{\omega}t}.$

Substituting this back into $z_i=v_i z_\eta(t) $, the common amplitude and phase of the $i^{\text{th}}$ oscillator in a cluster state can be obtained as
\begin{equation}
r_i = r_o = |v_i| \sqrt{\lambda + \eta_r \sigma \cos \beta - \eta_c \sigma \sin \beta}, 
\label{hohoho}
\end{equation}
and
\begin{align}
\phi_i = \omega t = \tilde{\omega} t + arg(v_i). 
\label{grr}
\end{align}
Equations \eqref{hohoho} and \eqref{grr} can be interpreted as
follows: For $|v_i|>0$, the $i^{\text{th}}$ node oscillates with a phase
shift of $arg(v_i)-arg(v_{i-1})$ with respect to the preceding node.
Thus, if all components $v_i$ are real and non-zero,  we obtain in-phase
oscillations. If $v_i=0$, the $i^{\text{th}}$ node undergoes amplitude
death. For eigenvectors containing both zero and non-zero elements we obtain
partial amplitude death.

\section{Linear Stability Analysis}
In the case of  in-phase and anti-phase
synchronization  a master-stability ansatz \cite{PEC98} is possible
(for details see \cite{POE15}). In \cite{POE15} a detailed stability analysis was given for in-phase
and anti-phase synchronization and partial amplitude-death
solution. In the following, we consider the stability of general
cluster states as given in Eqs.~\eqref{hohoho} and \eqref{grr}. 
For these patterns, an analytic solution is only possible in the
absence of  partial amplitude death. We will focus here on this
case. In the presence of partial amplitude death and cluster states,
numerical methods should be used.
%
Note that the analysis presented here is very similar to the one given in \cite{CHO09}.
We include it here to increase the comprehensiveness and
readability of our discussion.

We start by introducing polar coordinates $z_j=r_je^{\mathrm{i}\phi_j}$ in the system of coupled oscillators given by Eqs. (\ref{eq:f}) and \eqref{eq:ff} yielding
\begin{equation}
\begin{aligned}
 \dot{r}_i&=(\lambda-r_i^2)r_i+\sigma\sum_jA_{ij}r_j\cos(\beta+\phi_j-\phi_i)~,\\
 r_i\dot{\phi}_i&=\omega r_i+\sigma\sum_jA_{ij}r_j\sin(\beta+\phi_j-\phi_i)~.
\end{aligned}
 \label{eq:polareq}
\end{equation}
These equations need to be linearized around the solution given by
Eqs.~\eqref{hohoho} and \eqref{grr}. For $r_i=0$ the phase $\phi_i$ of
the oscillator is not defined. The use of polar coordinates therefore
is restricted to solutions without dead oscillators,
i.e. $|v_i|=1~\forall~i\in\{1,\dots,N\}$, corresponding to cluster
states, respectively. 

We introduce small perturbations, $\delta r_i$ and $\delta\phi_i$, of the limit cycle solution $z_i=v_iz_\eta$ in radius and phase
\begin{equation}
\begin{aligned}
 r_i=r_0(1+\delta r_i),\quad
 \phi_i=i\omega + \arg(v_i)  + \delta\phi_i
\end{aligned}
\label{eq:pertub}
\end{equation}
Inserting Eq. (\ref{eq:pertub}) into Eq. (\ref{eq:polareq}) and expanding all right hand sides around $\delta r_i=0,~\delta\phi_i=0$ up to first order yields
\begin{equation}
 \begin{aligned}
 \dot{\delta r_i}=& (\lambda-3r_0^2)\,\delta r_i+\sigma\sum_{j=1}^N A_{ij}\left[\cos\left(\beta+             arg(v_j)-arg(v_i)\right)\delta r_j\right.\\
&\left.-r_0\sin\left(\beta+arg(v_j)-arg(v_i)\right)\left(\delta\phi_j-\delta\phi_i\right)\right]~,\\[0.1cm]
\dot{\delta\phi_i}&=-\eta\sigma\sin\beta\delta r_i+\sigma\sum_{j=1}^NA_{ij}\left[\sin\left(\beta+arg(v_j)-arg(v_i)\right)\delta r_j\right.\\
&\left.+\cos\left(\beta+arg(v_j)-arg(v_i)\right)\left(\delta\phi_j-\delta\phi_i\right)\right]~.
 \end{aligned}
 \label{eq:bla}
\end{equation}

We write this in a compact vector from:
\begin{equation}
\dot \zeta =  (\mathbf{Q}  + \mathbf{R}) \zeta,
\label{zeta}
\end{equation}
where $\zeta= (\delta r_1,\delta\phi_1,\ldots,\delta r_N,\delta\phi_N)$.

$\mathbf{Q}$ is a block diagonal matrix where the $i$th block $\bold{Q}_i$ is
given by 
\begin{equation}
\bold{Q}_i=\begin{pmatrix} 
2r_0^2+\sum_{j=0}^NA_{ij}\cos\theta_j^i & -\sum_{j=0}^NA_{ij}\sin\theta_j^i \\
\sum_{j=0}^NA_{ij}\sin\theta_j^i & \sum_{j=0}^NA_{ij}\cos\theta_j^i 
\end{pmatrix}
\end{equation}

with $\theta_j^i = \beta + arg(v_j) - arg(v_i)$. $\mathbf{R}$ is also
a block matrix. $\bold{R}_{ij}$, the block on position $i$,$j$, reads
\begin{equation}
\bold{R}_{ij}=KA_{ij}\begin{pmatrix} 
\cos\theta_j^i & -\sin\theta_j^i \\
\sin\theta_j^i & \cos\theta_j^i 
\end{pmatrix}.
\end{equation}
In the case of in-phase and anti-phase
synchronization  a master-stability ansatz \cite{PEC98} is possible
(for details see \cite{POE15}), i.e., Eq.~\eqref{zeta} can be block
diagonalized. However, in the case of general cluster states as
considered here, this is not feasible because $\bold{Q}_i$ depends
on $i$. Instead, we calculate the Floquet multiplier $\mu$ directly 
as the eigenvalue of the matrix $\mathbf{Q}  + \mathbf{R}$. The
Floquet exponent $\Lambda$ can be obtained as  $\Lambda= \ln \mu$. The
real part of $\Lambda$ determines the stability of the considered
state: For $\text{Re}(\Lambda)<0$, the considered solution is stable,
for $\text{Re}(\Lambda)>0$ it is unstable.

  \section{Network topologies}
\label{top}
Here we present the three different methods of constructing the
hierarchical networks which are further studied in Sec.~\ref{results}.

\subsection{1D fractal}
\label{1D}
In this section, we elaborate on the method used to create the first
model of a hierarchical network (1D fractal network) in a ring topology. This model was
first introduced in \cite{OME15} to study chimera states. The network
is constructed by selecting a base pattern composed of ones and zeros. Then we
iterate over this base $n$ times, substituting the base pattern in each
iteration every time we encounter a one and a string of zeros of size
$b$ every time we come across a zero, where $b$ is the length of the
base. After $n-1$ iterations we obtain the $n^{th}$ hierarchy level. We then
have a string $S$ of size $b^{n}$. We use this
string of ones and zeros as the first row of the adjacency matrix. Each
following row of the matrix is obtained by shifting the previous row
by one element to the right applying periodic boundary
conditions. This results in  a circulant matrix. Circulant
matrices have well-known eigenvalues and eigenvectors, and are further
discussed in Sec.~\ref{results}\ref{iryna}.

\begin{figure}
  \noindent\stackinset{l}{-0.2cm}{t}{-0.6cm}{\colorbox{white}{(a)}}{
    \includegraphics[width=6cm, valign = t]{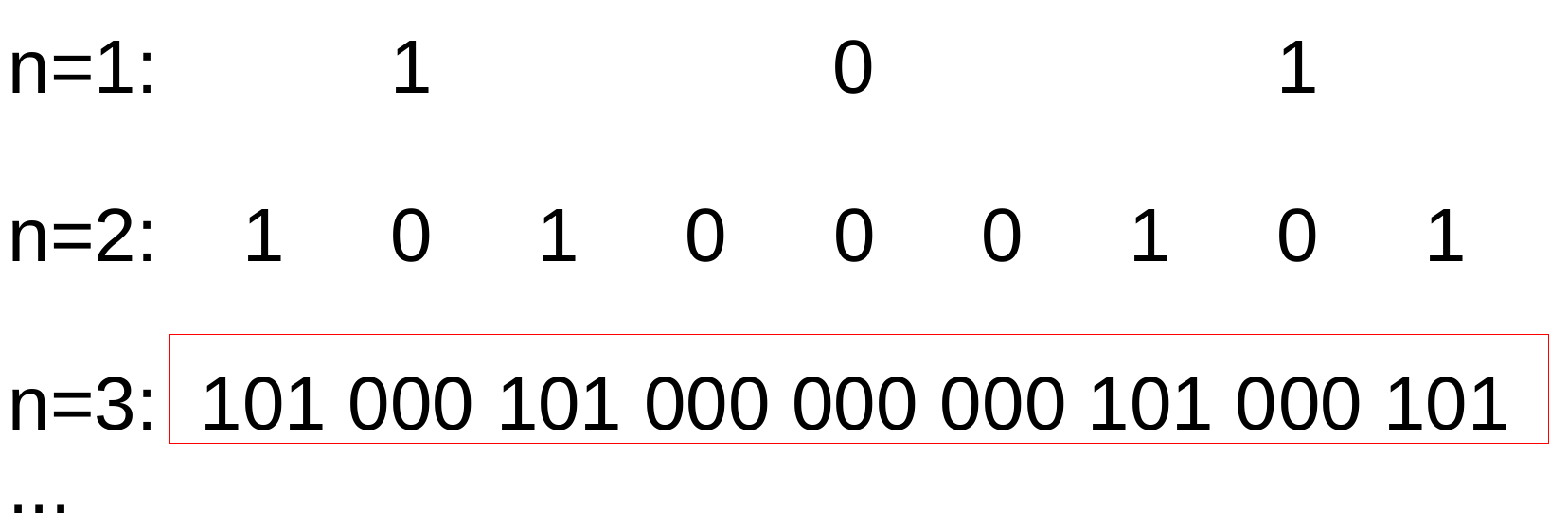}}        
  \hfill
  \centering
  \noindent\stackinset{l}{}{t}{-0.6cm}{\colorbox{white}{(b)}}{
    \includegraphics[width=6cm, valign = t]{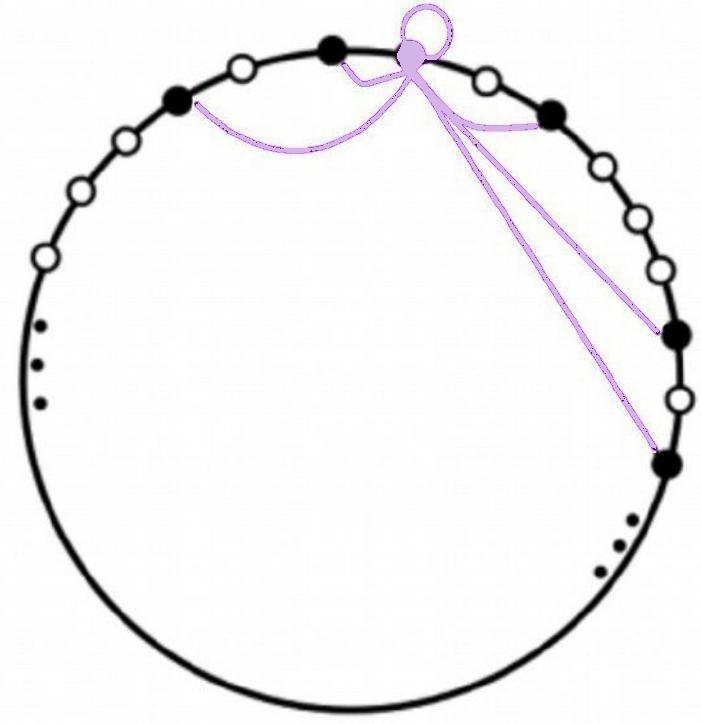}}
  \caption{Construction of model 1. (a) Iteration process
    for the construction of the hierarchical connectivity
    scheme for a
    triadic Cantor set (101) and two
    iterations.  (b) Schematic representation of the corresponding
    network topology.  Purple (gray) lines represent the links starting at the
    first node. For clarity, the links of the others nodes are not shown.}
  \label{hier1}
\end{figure}

As an example, let us consider a connectivity matrix generated from an
initiation string (base) of size $b = 3$ with a base pattern $(101)$. The base ($n=1$), the string after the first iteration ($n=2$),
and after the
second, in this case the final, iteration ($n=3$) are shown 
in Fig.~\ref{hier1}(a). The final string defines
the connections of the first node to the other nodes of the network. 
The
corresponding links are shown in purple (gray) in
Fig.~\ref{hier1}(b).
The connections of all other nodes follow the same rules (not shown in
Fig.~\ref{hier1}(b) for the sake of clarity), in other words each node
``sees'' the same network; this is equivalent to the circulant property
of the final network adjacency matrix. Thus, the adjacency matrix of our example is
given by
\begin{equation}
A  =
  \begin{pmatrix}
1&0&1& 0 & \ldots&  0 &1&0&1\\
1&1&0& 1 & \ldots&  0   &0&1&0\\
\vdots &\vdots &\vdots&\vdots&\vdots&\vdots&\vdots\\
1&0&0& 0&     \ldots& 0     &1&1&0\\
0&1&0&0& \ldots&  1&0&1&1\\
  \end{pmatrix}.
\end{equation}

\subsection{Modular fractal}

In this section, we extend the 1D fractal network presented in
Sec.~\ref{top}\ref{1D} to a 2D hierarchy. Here, instead of a base string, we use an initial $b \times b$ base matrix $A_1$ of ones and zeros. The $b^n \times b^n$  adjacency matrix $A$ for the
$n^{\text{th}}$ hierarchy level can then be formed by taking $n$ Kronecker
products of the initial adjacency matrix $A_1$ with itself, i.e., 
\begin{equation}\label{eq:4}
A= \overbrace{A_1 \otimes \ldots \otimes A_1}^\text{$n$ times}.
\end{equation}
This is essentially the 2D version of the procedure described in
Sec.~\ref{1D}: We start with a $b \times b$  base matrix $A_1$ of size $m \times m$. If we encounter a non-zero element
in the base, we substitute it with the element times the matrix $A_1$, whereas a zero is replaced
by a zero matrix of size equal to the size of $A_1$. 
We repeat this substitution procedure $n$ times resulting in the adjacency matrix $A$ of size $m^{n-1} \times m^{n-1}$
given in Eq.~\eqref{eq:4}. This matrix $A$ then defines the coupling topology, but it is no longer a circulant matrix, i.e., the ring topology is replaced by a modular topology. Note that by the method used to construct the adjacency matrix $A$, $A$ has a constant row sum if $A_1$ has a constant row sum. 

In order to be able to apply the eigensolution method (Sec.~\ref{top}), we choose the initial matrix $A_1$ to either be a circulant matrix (for example, the adjacency matrix from a network as described in Sec.~\ref{top}\ref{1D}), or an adjacency matrix whose eigenvector components satisfy Eq.~\ref{eq:1}, such as the mesoscale motifs studied in \cite{POE15}.

 \begin{figure}
 \centering
  \includegraphics[width=13cm, valign = t]{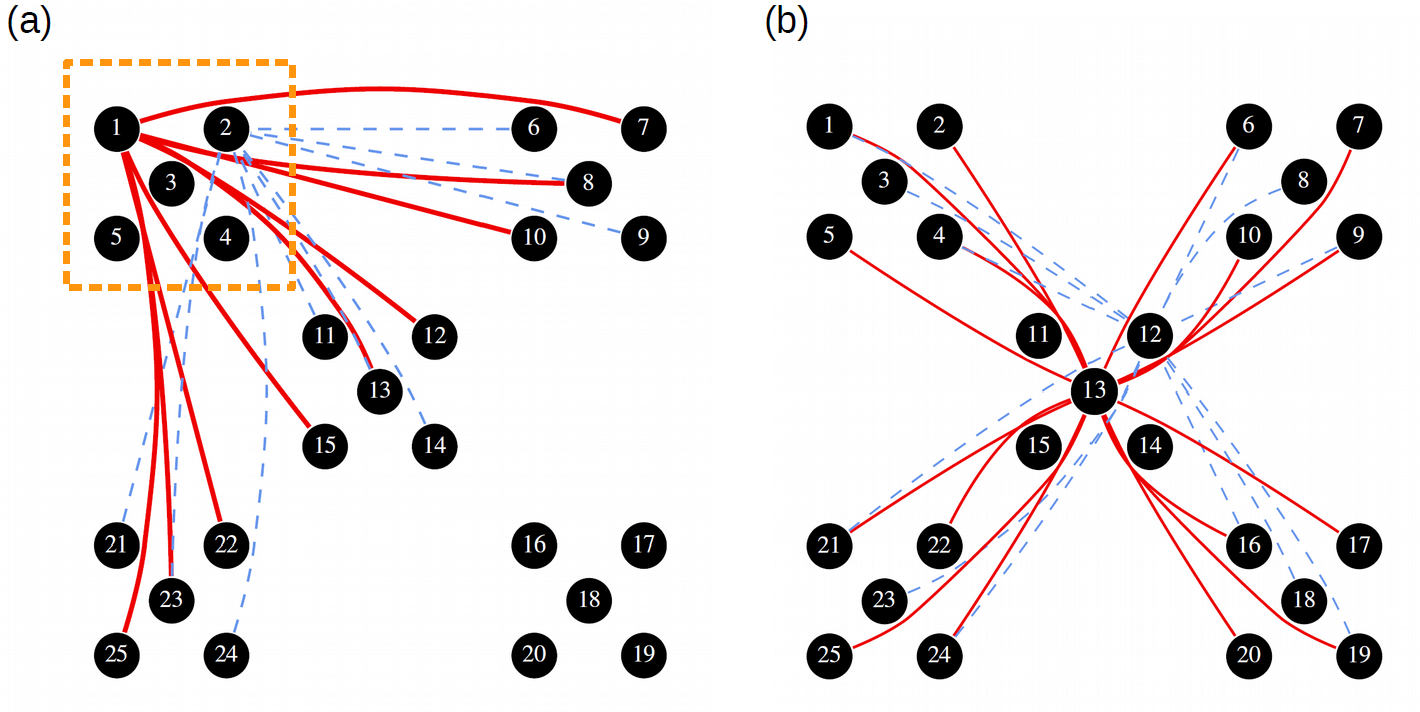}
 \caption{Schematic representation of model 2 for $n=2$ . (a) Red (black) solid lines and
   blue (gray) dashed lines represent the links starting at the
    first and second node, respectively. The 
initial  network motif given by
   the adjacency matrix $A_1$  is marked by the yellow (gray)
   dashed square. (b)  Red (black) solid lines and
   blue (gray) dashed lines represent the links starting at the
    $13^{th}$ and   $12^{th}$ node, respectively.
  The adjacency matrix is obtained by
 Eq.~\eqref{eq:4}. For clarity, the links of the others nodes are not
 shown in panels (a) and (b).}
 \label{plo}
 \end{figure}
 
As an example, let us consider the case of the five node motif shown in
Fig.~\ref{plo}(a) and investigated in \cite{POE15}. Figure~\ref{plo}(b) shows the network topology of the final adjacency matrix $A$ for $n=2$, i.e., $ A = A_1 \otimes A_1$. As one can see, the topology is not an intuitive extrapolation from 1D to 2D, however, it retains a fractal structure. 
The normalized adjacency matrix $A_1$ of this motif is given by
\begin{equation}
A_1 = 
\begin{pmatrix}
0 & \frac{1}{3} & \frac{1}{3} & 0 & \frac{1}{3} \\
\frac{1}{3} & 0 & \frac{1}{3}  & \frac{1}{3} & 0 \\
\frac{1}{4} & \frac{1}{4} & 0 & \frac{1}{4} & \frac{1}{4}\\
0 & \frac{1}{3} & \frac{1}{3} & 0 & \frac{1}{3} \\
\frac{1}{3} & 0 & \frac{1}{3} &  \frac{1}{3} & 0 \\
\end{pmatrix}.
\label{adjm}
\end{equation}

The resulting matrix $A$ is given explicitly in the Appendix A.3.

\subsection{ Hierarchical}
\label{modehier}

In this section, we present a hierarchical topology which is
self-similar on every scale. In \cite{DHU13}, synchronization with multiple delays has been investigated for this hierarchical network for the second level of the hierarchy ($n=2$). 

The network is created as follows: We start with a network motif given
by the $m \times m$ matrix $A_1 $. Then, the adjacency matrix $A$ after
one iteration ($n=2$) is given by 
\begin{equation}\label{eq:5}
A= A_1 \otimes E_m + \mathbf{1_m} \otimes A_1
\end{equation}
where $E_m$ is the $m \times m$ matrix with all entries equal to 1, and
$\mathbf{1_m}$ is the $m \times m$ identity matrix. $\mathbf{1_m} \otimes A_1$
represents the direct coupling inside the motifs, whereas $A_1 \otimes E_m$
is the mean field (all-to-all) coupling between the motifs.  The adjacency matrix for the $n^{\text{th}}$ level of the hierarchy  is
obtained by repeating the above procedure $n-1$ times  where we
replace $A_1$ in Eq.~\eqref{eq:5} by the normalized adjacency matrix of the previous iteration step. Note that by the method used here to construct $A$, $A$ has a constant row sum if $A_1$ has a constant row sum. 

As an example consider the case of the five node motif shown in Fig.~\ref{plo}(a) and investigated in \cite{POE15}. 
Its normalised adjacency matrix is given by Eq. \eqref{adjm}. Figure
\ref{plo}(a) depicts the final topology as calculated by
Eq.~\eqref{eq:5}. Clearly, the coupling between the motifs has the
same structure as the coupling between the nodes inside one motif giving
rise to a  self-similar architecture.
 
\begin{figure}
 \centering
  \includegraphics[width=0.5\textwidth]{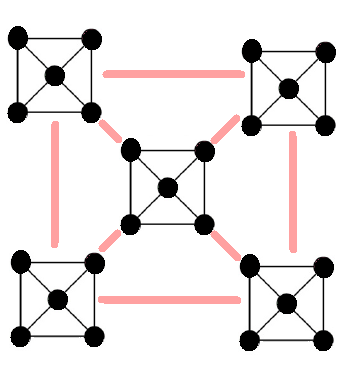}
 \caption{Schematic view of model 3 for the $n_1=2$ hierarchy  and  with the
   initial motif given by Eq.~\eqref{adjm} and depicted in Fig.~\ref{plo}(a). The red (gray) lines
   indicate the mean field or all-to-all coupling between motifs and the black lines
   mark the links inside the motif.}
 \label{plo2}
 \end{figure}

\section{Results}
\label{results}
In this section, we study the  dynamics for the network models discussed in
Sec.~\ref{top}. In particular, we discuss the application of the eigensolution method
to hierarchical networks. We support our analytical
results with numerical simulations.
 \subsection{1D fractal}
\label{iryna}
This network has a circulant
adjacency matrix, which has been introduced in Sec.~\ref{analytic}. The general form of a circulant matrix is given by

\[
C=
\begin{bmatrix}
c_1     & c_{N} & \dots  & c_{3} & c_{2}  \\
c_{2} & c_1    & c_{N} &         & c_{3}  \\
\vdots  & c_{2}& c_1    & \ddots  & \vdots   \\
c_{N-1}  &        & \ddots & \ddots  & c_{N}   \\
c_{N}  & c_{N-1} & \dots  & c_{2} & c_1 \\
\end{bmatrix}.
\]

Its normalized eigenvectors \cite{GRA05} are given by 

\begin{equation}
\label{eigen}
v^j=\frac{1}{\sqrt{N}} (1,~ \omega_j,~ \omega_j^2,~ \ldots,~ \omega_j^{n-1})^T,\quad j= 1,\ldots, N,
\end{equation}
where $\omega = \exp(\frac{2 \pi \mathrm{i} j}{N})$.
The corresponding eigenvalues are given by
\begin{equation} 
\eta_j = c_1 + c_{N} \omega_j + c_{N-1} \omega_j^2 + \ldots + c_{2} \omega_j^{n-1}, \qquad j=1 \ldots N.
\end{equation}

From Eq.~\eqref{eigen} it follows that the eigenvectors of circulant
matrices have components which are given by the
roots of unity. Hence, they fulfil the property required by the eigensolution approach
presented in Sec.~\ref{analytic}, i.e., $|v_i| \in \{0,1 \}$, where we
defined $v_i$ as the $i^{\text{th}}$ component of the eigenvector $v$.  

In accordance with the eigensolution
method, substituting Eq. \eqref{eigen} into Eq. \eqref{grr} yields the $j^{\text{th}}$ eigensolution for the $k^{\text{th}}$ oscillator, $ k = 1, \ldots, N$,
\begin{equation}
z_k=v_k z_\eta(t) = \exp\left(\frac{2 \pi\mathrm{i}  j}{N}\right)^k
\sqrt{\tilde{\lambda}} \exp\left(\mathrm{i}\tilde{\omega} t\right) =
\sqrt{\tilde{\lambda}} \exp\left[\mathrm{i}\tilde{\omega} t + \frac{2 \pi \mathrm{i} k j}{N}\right].
\label{osc}
\end{equation}

From Eq.~\eqref{osc} it follows that the $j^{\text{th}}$ eigensolution is a
cluster state with a constant phase shift of $2 \pi j/N$ between
neighboring nodes. As discussed in Sec.~\ref{analytic}, the number of
clusters is then given by $lcm(j,N)/j$ \cite{CHO09}. The
eigenvectors of circulant matrices as given by Eq.~\eqref{eigen}  do not have
components equal to zero, and hence for non-degenerate eigenvalues, we do not find
partial amplitude death eigensolutions in this hierarchical
topology. For degenerate eigenvalues, we can obtain dead nodes by
linear combinations of the corresponding eigenvectors. This is further
discussed in the next subsection.
\subsubsection{Partial amplitude death in hierarchical networks} 
\label{wooho}

For model (a), partial amplitude death  eigensolutions do not exist for
non-degenerate eigenvalues. However, if the adjacency matrix of the
network has degenerate eigenvalues ($ \eta_{j_1} = \eta_{j_2}$ or $
\eta_{j_1} = \eta_{j_2}^*$ where 
$ j_1,j_2 \in \{1, \ldots, N\}$), we can find a new basis
$\tilde{v}^{j_1}$ and $\tilde{v}^{j_2}$ of the subspace spanned by the
corresponding eigenvectors $v^{j_1}$ and $v^{j_2}$ such that at least
one of the entries of $\tilde{v}^{j_1}$ or $\tilde{v}^{j_2}$ is 0.  We
start by selecting a position $m^*$, $m^* \in \{1, \ldots, N\}$, in the
eigenvector which we wish to convert to zero.
We achieve this by constructing $\tilde{v}^{j_1}$ as
\begin{equation}
\tilde{v}^{j_1} = v^{j_1} l - v^{j_2}
\label{cond2}
\end{equation}
where $l$ is given by
\begin{align} \label{l}
l=\frac{v^{j_2}_{m^*}}{v^{j_1}_{m^*}}
\end{align}
and $v^{j_i}_{m^*}$ denotes the $(m^*)^{\text{th}}$ element of the vector $v^{j_i},
i \in 1,2 \ldots N $. For a circulant adjacency matrix, Eq.~\eqref{l} reads
\begin{equation}
l=\exp{\left[2 \pi \mathrm{i} (j_2-j_1) {m^*} / N\right]} .
\label{koko}
\end{equation}
 Let $B$ be the set of integers $m \in \{1,\ldots, N\}$ for which the
 Eq. \eqref{cond2} holds true. We require that the set $B$ always has at
 least one element, namely ${m^*}$, the position we initially chose to
 be zero. Trivially, the cardinality of $B$ equals the number of zeros in $\tilde{v}^{j_1}$.

To ensure the existence of the eigensolution, we require that the original assumption made while finding the eigensolutions still holds for the new eigenvector, i.e., the following condition is satisfied:
\begin{equation}
| \tilde{v}^{j_1}_i |^2 \in \{0,1\} 
\label{cond1]}
\end{equation}  
for $ \forall i \in 1 \ldots N$.
 
While this condition is fulfilled for the eigenvectors calculated
according to Eq. \eqref{eigen}, it is not automatically fulfilled for
a linear combination of these eigenvectors. We can rewrite the
conditions for the existence of an eigensolution by substituting Eqs. \eqref{eigen} and \eqref{cond2} into Eq. \eqref{cond1]}:
\begin{equation}
|\exp{\left[2 \pi \mathrm{i} (j_2) k / N\right]} - l \exp{\left[2 \pi \mathrm{i} (j_1) k / N \right]}|^2 = c.
\label{zozo}
\end{equation}
for $ \forall k \in \{ 1, \ldots, N\} $ such that $k \notin B $, i.e., for all nonzero entries of the eigenvector $\tilde{v}^{j_1}$.The normalization constant is $c$.
Substituting for $l$ from Eq. \eqref{koko} into Eq. \eqref{zozo}, we obtain
\begin{equation}
\exp\left[\frac{2 \pi \mathrm{i} (j_2-j_1) ({m^*}-k)} {N}\right] = c.
\label{three}
\end{equation}
If this condition holds $\forall k \notin B$, an eigensolution exists
for the corresponding values of $j_1$, $j_2$, which represents a state
where all nodes $m\in B$ are amplitude dead.

\begin{figure}
    \centering
    \noindent\stackinset{l}{}{t}{-0.6cm}{\colorbox{white}{(a)}}{
        \includegraphics[width=5cm, valign = t]{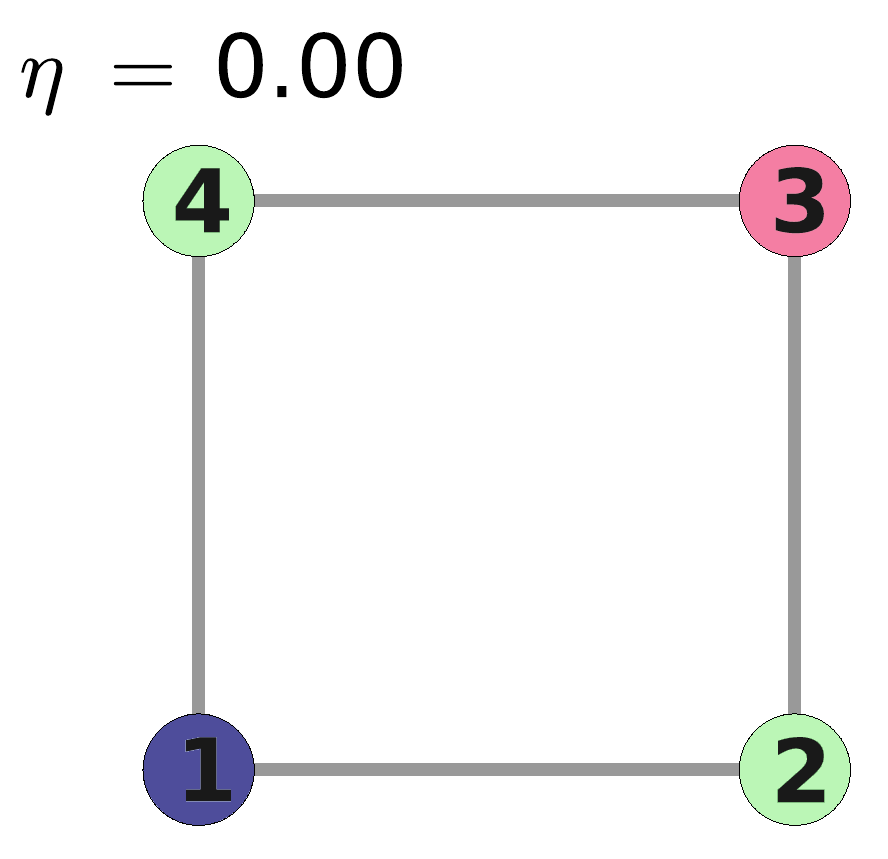}}
    \hfill
    \centering
    \noindent\stackinset{l}{}{t}{-0.6cm}{\colorbox{white}{(b)}}{
        \includegraphics[width=7cm, valign =t]{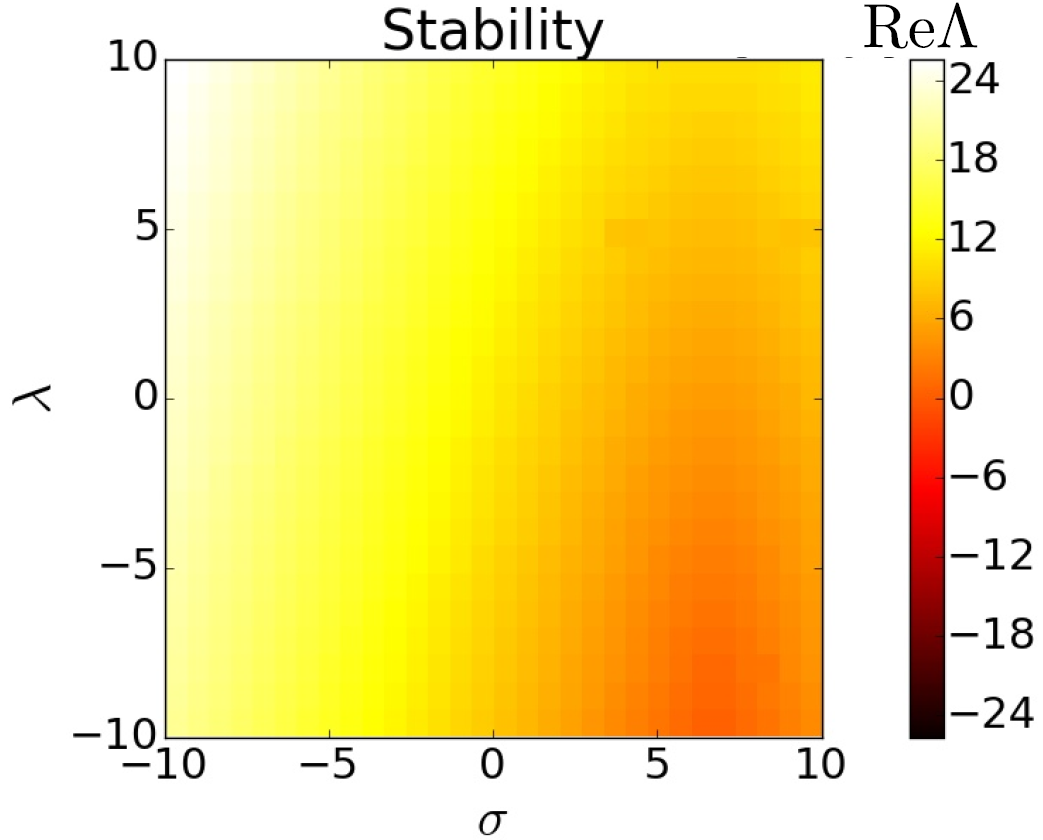}}
        \caption{Example of partial amplitude death. (a) Four node
          network motif colored according to the predicted
          eigensolution. Green (light grey) marks nodes which exhibit amplitude
          death. The red (intermediate grey) and the blue node (dark
          grey) exhibit antiphase-synchronized oscillations with
          respect to each other. (b) Largest real part of the Floquet
          exponents $\Lambda$
          for the motif and the eigensolution shown in
          (a).}
\label{floq}
    \end{figure}

As an example, we consider the four-node ring network shown in
Fig.~\ref{floq}(a) and investigated in \cite{POE15}. For this network,
$\eta_1=\eta_3=0$ holds. The corresponding normalized eigenvectors are
 $v^1= 0.5\{1, i,-1, -i\}$  and $v^3=0.5 \{1, -i,-1, i\}$. For
 ${m^*}=2$, $l$ is calculated to be $-1$ and $\tilde{v}^{j_1} = \{ 1,
 0,-1, 0 \} $. The set of all indices for which $\tilde{v}^{j_1}_i=0$
 is $B= \{ 2,4 \}$. For all $k \notin B$, i.e., $k=1, 3$,
 Eq. \eqref{three} is satisfied. Therefore, a solution of the form
 $\tilde{v}^{j_1}$ exists, where nodes 1 and 3 are antiphase-synchronized
 with respect to each other (marked by blue and
red color in Fig.~\ref{floq}(a)), and nodes 2 and 4 are amplitude
dead (the green nodes in Fig.~\ref{floq}(a)). 

The stability of the solution is found by using the linear stability analysis for partial amplitude death states suggested in \cite{POE15}. The value of the real part of the largest Floquet exponent for this network is positive throughout the parameter space (as seen in Fig.~\ref{floq}(b)). Numerically, we have investigated networks up to 800 nodes, and all partial amplitude death solutions we have found are unstable.

\subsubsection{Oscillation death}    

In this section, we show that oscillation death can arise as  a cluster
state with vanishing  common frequency, even for symmetric coupling. 
For the sake of
simplicity, let us consider a network with base $(0110)$ and $n=2$,
i.e., $N=16$. We investigate  the $j=4$-cluster state with number of
clusters given by $M= lcm(4,16)/4 = 4$.

\begin{figure}
  \centering
  \noindent\stackinset{l}{}{t}{-0.6cm}{\colorbox{white}{(a)}}{
    \includegraphics[width=8cm]{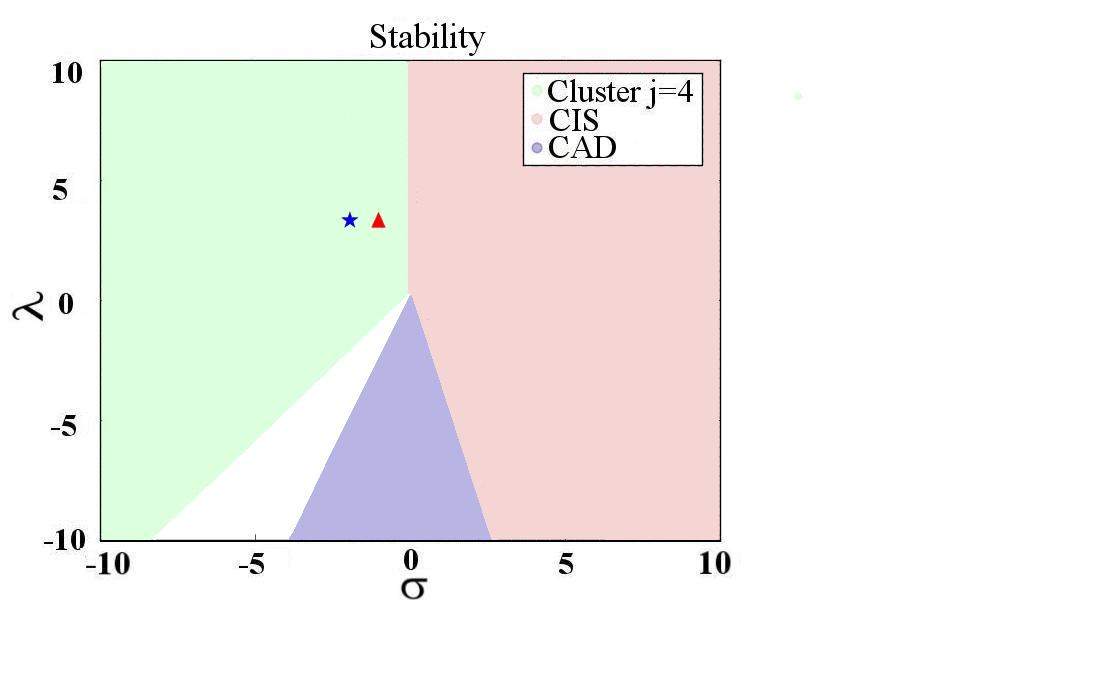}}
  \vfill
  \centering
  \noindent\stackinset{l}{}{t}{-0.6cm}{\colorbox{white}{(b)}}{
    \includegraphics[width=5cm]{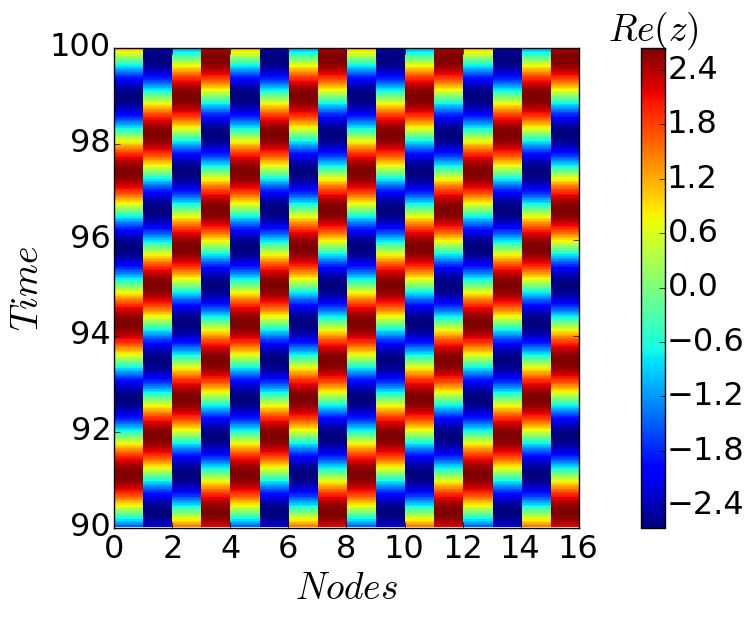}}
  \hspace{1mm}
  \centering
  \noindent\stackinset{l}{}{t}{-1.6cm}{\colorbox{white}{(c)}}{
    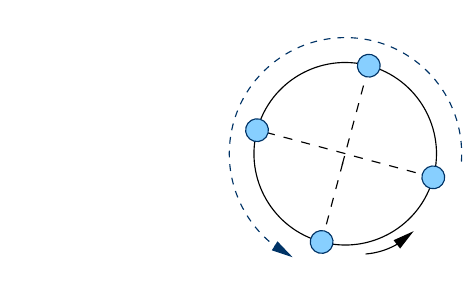}
  \vfill
  \noindent\stackinset{l}{}{t}{-0.6cm}{\colorbox{white}{(d)}}{
    \includegraphics[width=5cm]{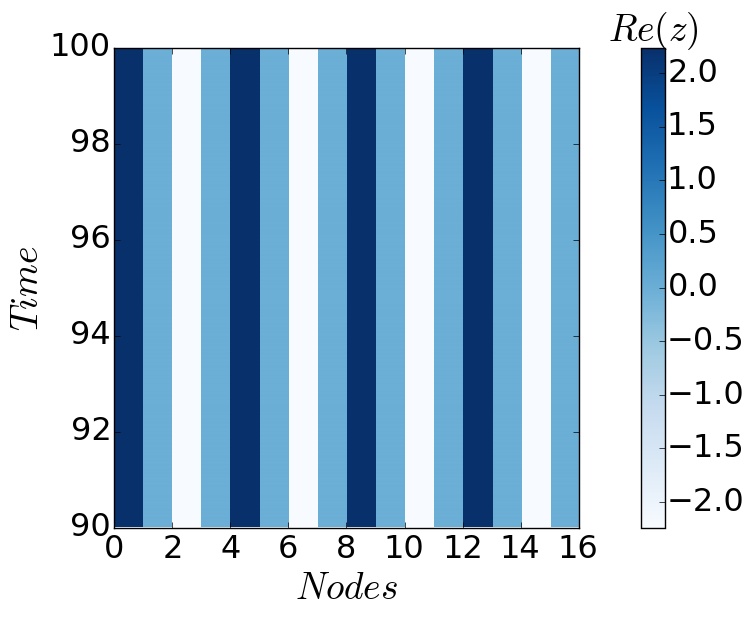}}
  \hspace{1mm}
  \centering
  \noindent\stackinset{l}{}{t}{-1.6cm}{\colorbox{white}{(e)}}{
    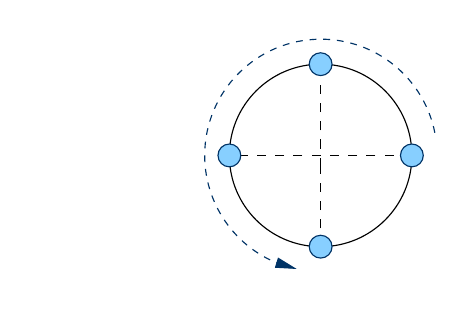}
  \hfill
  \caption{Eigensolutions of the 1D fractal model.
    (a) Linear stability analysis. CIS stands for complete in-phase
    synchronization. Parameter values where this state is stable are marked by red (intermediate gray) shading. Blue 
    (dark gray) shading denotes the stability region of the complete
    amplitude death (CAD) state, i.e., a state where all nodes are in the trivial steady state.
    Green (light gray) shading denotes the stable region of a $j=4$ cluster state.
    (b) Space-time plot for $\sigma =-2$ (marked by the blue star in panel (a)). The corresponding state is
    schematically depicted in (c).
    (d) Space-time plot for $\sigma =  -1$ (marked by the red triangle in panel (a)). The corresponding state is
    schematically depicted in (e).  Other parameters:  base $(0110)$, $n=2$,
    $N=16$, $j=4$, $\lambda =3$, $\omega =2$, $\beta =0$.}%
  \label{fig:halo}%
\end{figure}

For the considered network, Fig.~\ref{fig:halo} depicts the stable
regions for the  $4$-cluster state in green (light gray), for
in-phase synchrony in red (intermediate gray), and for complete
amplitude death in blue (dark gray). In the white region we obtain
solutions that do not fall into any of the categories we study,
i.e.,  in-phase synchronization, cluster synchronization,  or partial amplitude 
death. The solutions in the white region include patterns with non-equal
radii and/or non-equal frequencies.
For $\lambda = 3, \sigma = -2$, a
 $4$-cluster state is observed, as predicted by the stability plot in
Fig.~\ref{fig:halo}(a) (corresponding parameter values are marked by a green
star). The corresponding space-time plots are shown in
Fig.~\ref{fig:halo}(b), respectively, and show that the
cluster state has a nonvanishing frequency, i.e, $\tilde \omega \neq
0$.  For a schematic representation of this state see Fig.~\ref{fig:halo}(c). For $\lambda = 3, \sigma = -1$ (marked by a red triangle in
Fig.~\ref{fig:halo}(a)), we also observe that the $4$-cluster state is stable, however, here
the common frequency as given by Eq.~\eqref{hohoho}, is zero as can be seen in the corresponding space-time plot
(Fig.~\ref{fig:halo}(d)). Thus, for these parameters we obtain oscillation death. A schematic figure of this states is
shown in Fig.~\ref{fig:halo}(e).

In addition, it is also possible to obtain a mixed-death state
consisting of coexisting partial oscillation death and partial
amplitude death. Oscillation death occurs as a consequence of
vanishing oscillation frequency for pairs of nodes whose corresponding
eigenvector components are non-zero and symmetric about zero, whereas
the eigenvector component corresponding to nodes that are in the
amplitude death state is exactly zero. This state can be found for
motifs in \cite{POE15}: We consider a motif or network showing
amplitude death, e.g., the motif shown in Fig. \ref{floq}(a), and vary
the coupling strength such that the frequency of the oscillating
nodes becomes zero, i.e., we change $\sigma$ in Eq.~\eqref{wtilde}
such that $\tilde \omega=0$. This means that the oscillating nodes
stop oscillating with a non-zero radius given by $r_0$ (see
Eq.~\eqref{wtilde}), while the nodes which where amplitude-dead from
the beginning remain at the origin (recall that in Eq.~\eqref{wtilde}, 
$|v_0|=0$ corresponds to the nodes undergoing amplitude death). Note
that the stability analysis for oscillation death is the same as
  that for cluster states, since oscillation death arises  here as a cluster state with $\tilde{\omega}=0$.



\subsection{Modular fractal}
\label{2dhier}
Recall that the adjacency matrix for the $n^{\text{th}}$ level of the
hierarchy is given by $A = \overbrace{ A_1 \otimes \ldots \otimes
  A_1}^\text{$n$ times}$, where $A_1$ is our initial motif. The
eigenvectors of $A$ are then given by the Kronecker product of
combinations of $n$ eigenvectors of $A_1$,  i.e.,  by 
\begin{equation}
v=\overbrace{v^i \otimes v^j \ldots \otimes v^k }^\text{$n$ times} \quad \quad i,j,k \in \{1, \ldots m \}
\label{eigvecasd}
\end{equation} 
where $v^i ,v^j \ldots , v^k$ are chosen from the eigenvectors of $A_1$. Refer to Sec.~\ref{app1} in Appendix $A$ for the proof.

\begin{figure}
 \centering
 \noindent\stackinset{l}{}{t}{-0.6cm}{\colorbox{white}{(a)}}{
 \includegraphics[width=3cm, valign = t]{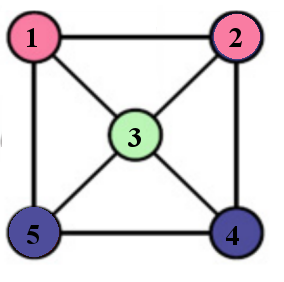}}
 \hfill
 \noindent\stackinset{l}{}{t}{-0.6cm}{\colorbox{white}{(b)}}{
  \includegraphics[width=0.5\textwidth, valign = t]{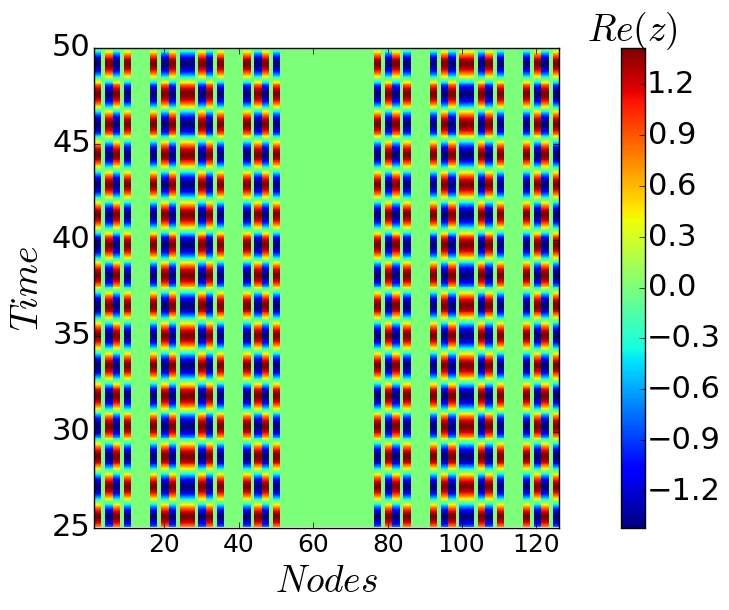}}
 \caption{Partial amplitude death in the modular fractal model and the hierarchical model. 
(a) Initial network motif color-coded according to the chosen
   eigenvector (1,1,0,-1-1). The green node shows amplitude death, the
   red and blue nodes are antiphase-synchronized with respect to each
   other. (b) space-time plot for the $n=3$ level of the hierarchy for the network motif in
   (a). Parameters: $\sigma= -5, \lambda=1, \omega =2, \beta=0.$}
 \label{plko}
 \end{figure}

Here, we consider the five-node motif shown in Fig.~\ref{plo}(a), for
$n=3$ and the eigenvector $v \otimes v
\otimes v$, where $v= (1,1,0,-1,-1)$. As seen in Fig.~\ref{plko}, the
network dynamics is nested or self-similar on every scale because we choose $i=j=k$ in Eq. \ref{eigvecasd}. In other words, since we choose the same eigenvector three times, we obtain a three-fold hierarchy in the dynamics. On the largest scale, the
corresponding nodes of the first two groups of 25 nodes each behave in
an identical fashion, and are antiphase-synchronized with the corresponding
nodes of the last two groups of 25 nodes. The middle group is
amplitude dead. Within each group, the first two motifs are in
synchrony with respect to each other, while the last two are in
antiphase-synchronization with respect to the first two, and the middle motif is dead. And finally,
within a motif, the dynamics follows the same pattern, i.e., the first
two nodes are antiphase-synchronized with respect to the last two nodes,
and the middle node is dead. Schematically, the dynamics of the motif
is shown in  Fig.~\ref{plko}(a).

If we choose the initial matrix $A_1$ to be a circulant
matrix obtained from a hierarchical network as created in
Sec.~\ref{results}\ref{iryna}, we obtain a 2D modular hierarchical structure: The initial
$b^{n_1} \times b^{n_1}$ matrix $A_1$ has a fractal topology with a repeating base, and the
final matrix $A$ has the same coupling structure with respect to these modules $A_1$. 
Thus we have two parameters that determine the hierarchy: $n_1$ that decides the level of hierarchy in $A_1$, 
and $n$ that sets the level of multi-hierarchy in $A$.

The $n^{\text{th}}$ level of the hierarchy can then be created by taking the n-fold Kronecker product of this matrix with itself.  Thus, the total number of nodes are $N = b^{n_1 \cdot n}$. The adjacency matrix $A$, when $A_1$ is circulant, is a block circulant matrix \cite{TEE08}.

Let us consider first the case $n=2$. We choose the
$o^{\text{th}}$ and the $p^{\text{th}}$  eigenvector of our initial $m\times m$ matrix $A_1$.
According to Eq. \eqref{eigen} and Eq. \eqref{eigvecasd},
the components of the final eigenvector are then given by:
\begin{equation}
 v_{j \cdot m+k}= \left(v^o \otimes v^p\right)_{j \cdot m+k}= \exp \left( \frac{2 \pi \mathrm{i} j o}{m}\right) \cdot \exp \left( \frac{2
   \pi \mathrm{i} k p}{m}\right) = \exp \left( \frac{2 \pi \mathrm{i} \left(jo+kp\right)}{m}\right) 
 \label{clust}
 \end{equation}
where $ k,j \in \{ 1, \ldots, m \} $. Thus, by the eigensolution
method, the nodes in each group are in a cluster state with a constant
phase shift of $\exp ( 2\pi\mathrm{i} p/m)$ between neighboring nodes, and the
corresponding nodes of the groups are in a cluster state with a
constant phase shift of $\exp ( 2\pi\mathrm{i} o/m)$ between neighboring
groups. Hence, we obtain hierarchical dynamics: a cluster state of
cluster states.

The components of the final eigenvector for $n=3$ are then given by
\begin{eqnarray}
v_{j\cdot m^2 + k\cdot m + l} &=&
\left(v^o \otimes v^p\otimes v^q\right)_{j\cdot m^2 + k\cdot m + l}=
 \exp \left( \frac{2 \pi \mathrm{i} jo}{m}\right) \exp \left( \frac{2 \pi \mathrm{i}
     kp}{m}\right) \exp \left( \frac{2 \pi  \mathrm{i}lq}{m}\right) \nonumber \\
&=& \exp \left( \frac{2 \pi \mathrm{i}\left(jo+kp+lq\right)}{m}\right),
 \label{clust11}
 \end{eqnarray}
where we use the $o^{\text{th}}$, $p^{\text{th}}$, and $q^{\text{th}}$  eigenvector of $A_1$.

As an example, we now consider two networks with the same base $(101)$
but with interchanged values of $n, n_1$ to demonstrate the direct
dependence of the dynamics on the chosen hierarchy and multi-hierarchy. 

\begin{figure}
  \centering
  \noindent\stackinset{l}{}{t}{-0.6cm}{\colorbox{white}{(a)}}{
    \includegraphics[width=4cm, height=5cm]{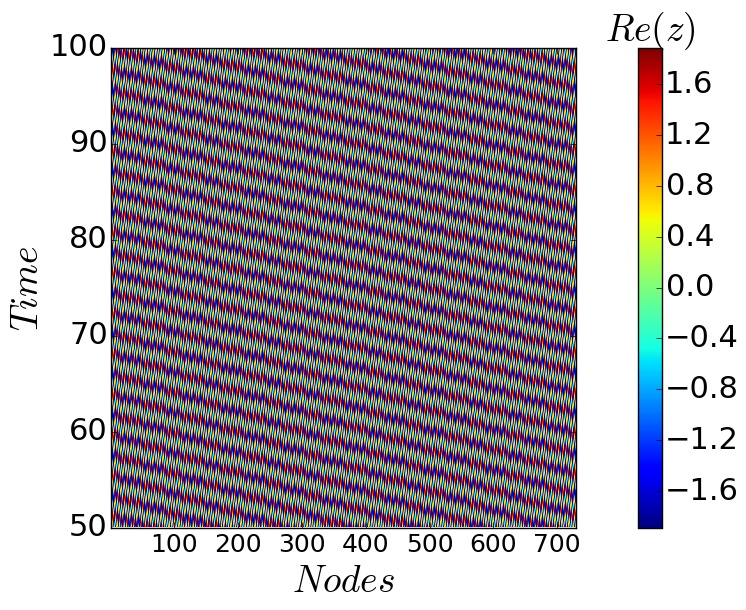}}
  \hfill
  \centering
  \noindent\stackinset{l}{}{t}{-0.6cm}{\colorbox{white}{(b)}}{
    \includegraphics[width=4cm, height=5cm]{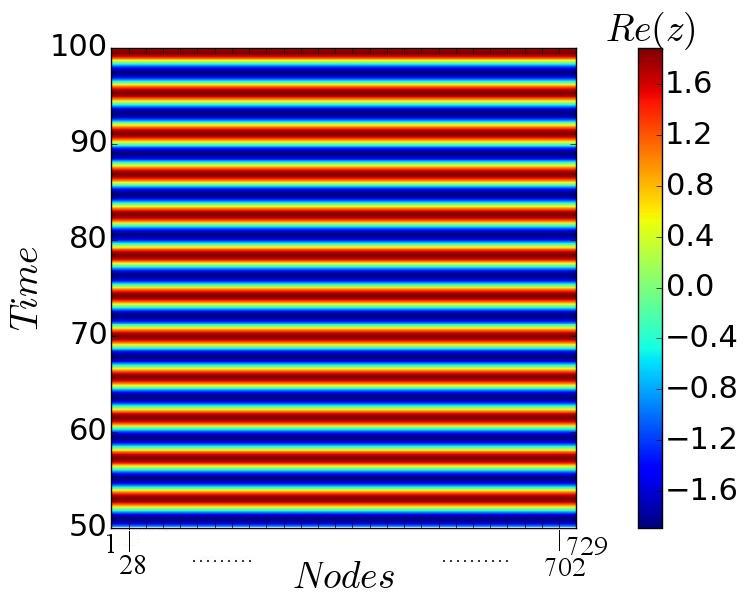}}
  \hfill
  \centering
  \noindent\stackinset{l}{}{t}{-0.6cm}{\colorbox{white}{(c)}}{
    \includegraphics[width=4cm, height=5cm]{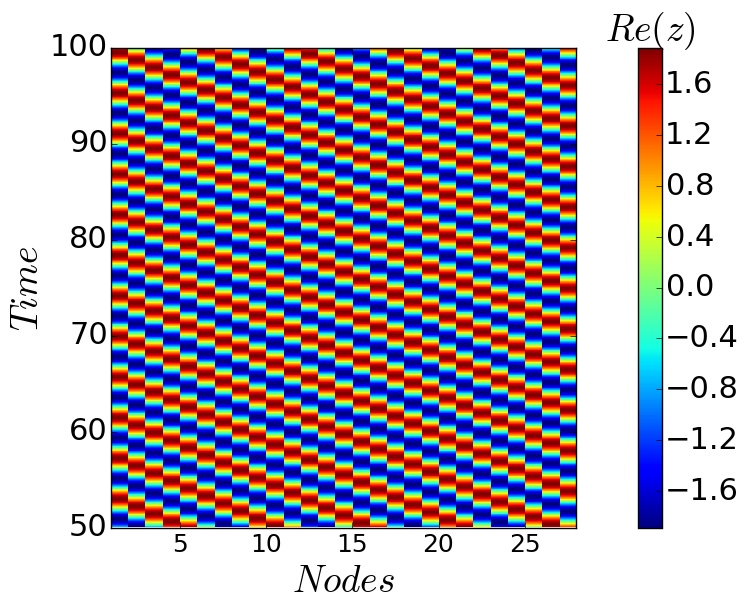}} 
  \caption{Hierachical dynamics in the modular fractal model for  $n_1=3$, $n=2$, and base $(101)$. 
	(a) Space-time plot,  (b) Space-time plot for every $27^{\text{th}}$
    node (first node of every group on the largest scale), (c) Space-time plot for the first group of 27 nodes. Other parameters: $ \sigma = -5/64, \beta=0, \lambda=1, \omega=2$.}
  \label{101_n2n13}
\end{figure}

In Fig.~\ref{101_n2n13}, we use the base $(101)$ with $n_1=3$, $n=2$
respectively. There are $N=(3^3)^2 = 729$ nodes. We chose to consider the state
corresponding to $o=0$ and $p=17$ in Eq.~\eqref{clust}.
Since $n=2$, we have a two-hierarchy cluster state, with 27 groups
of 27 nodes each: Corresponding nodes in each group (every $27^{\text{th}}$
node) are fully synchronized (Fig.~\ref{101_n2n13} (b)) because
$o=0$. Consecutive nodes in each group are phase shifted with respect
to each other with a constant phase of $ \phi = 2 \pi  17 / 27 $
as a result of $p=17$  (Fig.~\ref{101_n2n13} (c)).

\begin{figure}
  \centering
  \noindent\stackinset{l}{}{t}{-0.6cm}{\colorbox{white}{(a)}}{
    \includegraphics[width=13cm]{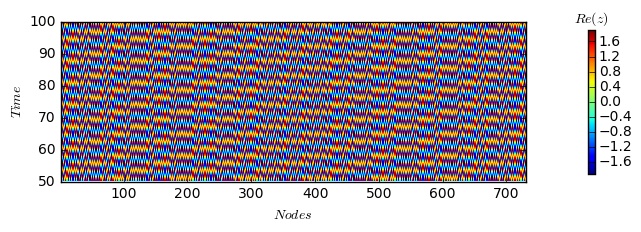}}
  \vfill
  \centering
  \noindent\stackinset{l}{}{t}{-0.6cm}{\colorbox{white}{(b)}}{
    \includegraphics[width=4cm, height=5cm]{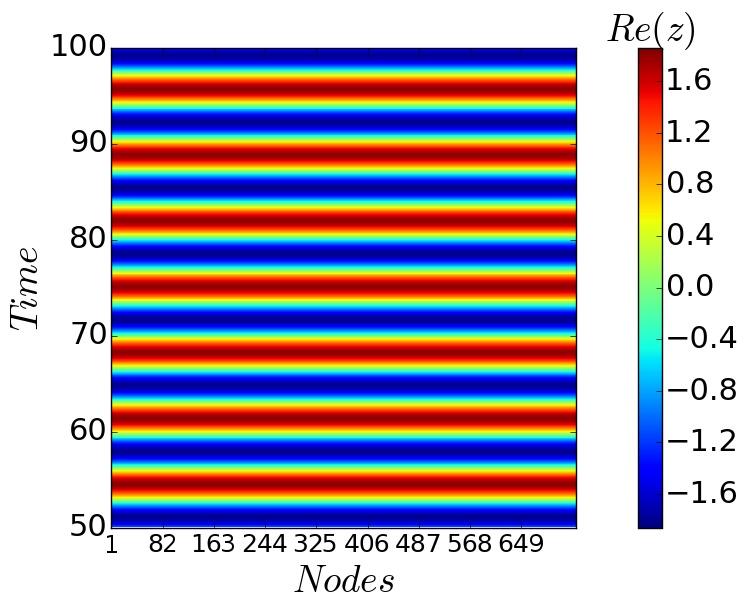}} 
  \hfill
  \centering
  \noindent\stackinset{l}{}{t}{-0.6cm}{\colorbox{white}{(c)}}{
    \includegraphics[width=4cm, height=5cm]{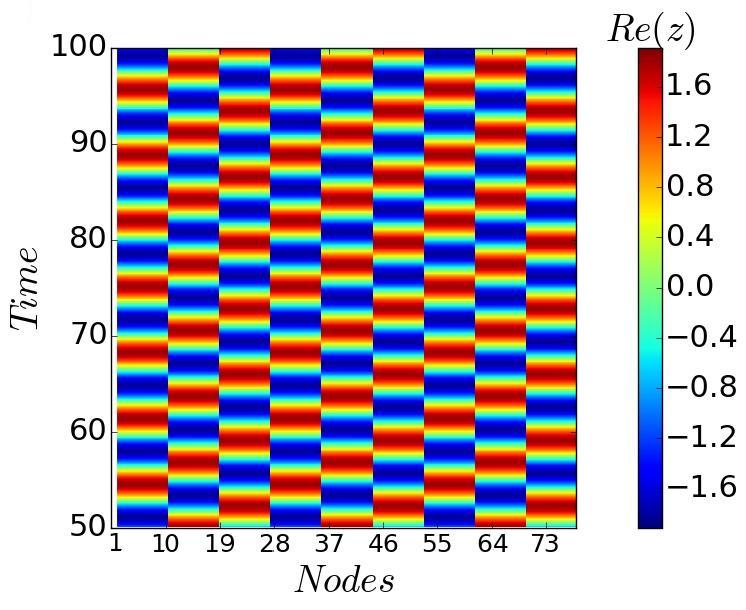}} 
  \hfill
  \centering
  \noindent\stackinset{l}{}{t}{-0.6cm}{\colorbox{white}{(d)}}{
    \includegraphics[width=4cm, height=5cm]{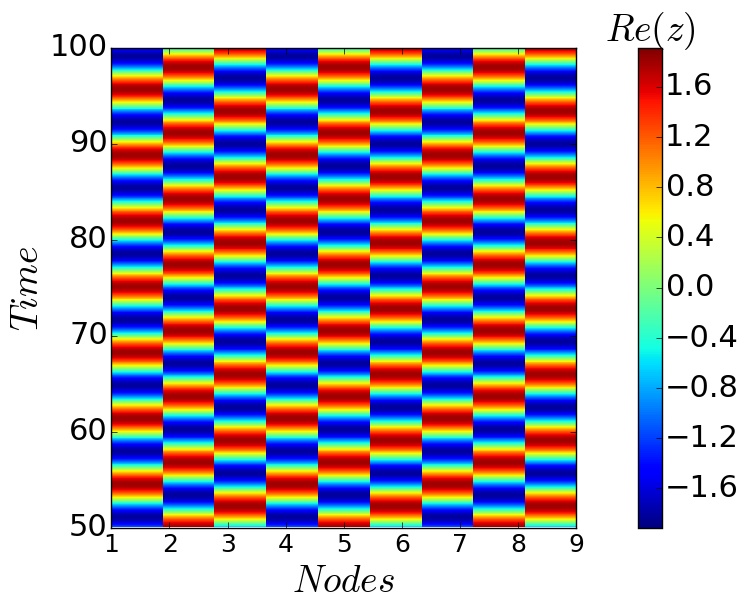}} 
  \caption{Hierachical dynamics in the modular fractal model for  $n_1=2$, $n=3$, and base $(101)$. 
	(a) Space-time plot. (b) Space-time plot for every $81^{st}$ node (first node of each group on the largest scale). (c) Space-time plot for every  of $9^{\text{th}}$ node in the first group (first node of each subgroup). (d) Space-time plot for the 9 nodes in the first subgroup (lowest scale). Other parameters: $ \sigma = -5/64, \beta=0, \lambda=1, \omega=2$.}
  \label{101_n3n12}
\end{figure}

In Fig.~\ref{101_n3n12}, we consider the base $(101)$ with $n_1=2$ and $n=3$. Since $n=3$, we anticipate a three-level hierarchy of cluster states: cluster state of cluster states of cluster
states.  We choose $o=0$, $p=5$, and $q=8$ in
Eq.~\eqref{clust11}. There are $N=(3^2)^3 = 729$ nodes with 9 groups
of 81 nodes each. Each group consists of 9 subgroups of 9 nodes
each. On the smallest scale, nodes in each subgroup are in a splay
state, i.e., neighboring nodes are phase shifted with respect to each
other with a constant phase of $ \phi = 2 \pi 8 / 9$
(Fig.~\ref{101_n3n12}(d)) because $q=8$.  Each subgroup within a group is also in a
cluster state, i.e., every $9^{\text{th}}$ node has a constant
phase shift of $ \phi = 2 \pi \cdot 5 / 9$ with respect to the previous
(Fig.~\ref{101_n3n12}(c)), this is due to $p=5$. Finally, on the large scale, every group is
synchronized meaning that  the phase lag between every
$81^{\text{st}}$ node is zero (Fig.~\ref{101_n3n12}(b)) corresponding to $o=0$. 
In summary,  we observe that $n=2 $ shows a two-level hierarchy in the
dynamics, whereas $n=3$ shows a three-level hierarchy. Analogously,
for  an $n$-level multi-hierarchy of the adjacency matrix, the dynamics is given by $n$ nested cluster states.

\subsection{Hierarchical}
In this model, the topology is self-similar on each scale. The
creation of this network is elaborated upon in Sec.~\ref{top}{\ref{modehier}. The
eigenvector of the adjacency matrix $A$ of the $n^{\text{th}}$ hierarchy can be
written as $n$ Kronecker products of the original eigenvectors of
initial the motif $A_1$, i.e.,
\begin{equation}
v=\overbrace{v^i \otimes v^j \ldots \otimes v^k }^\text{$n$ times} \quad \quad i,j,k \in \{1, \ldots m \}
\label{eigvecyxc}
\end{equation}
where $v^i,v^j, \ldots, v^k$ are chosen from the set of eigenvectors of
$A_1$.  However, in contrast to the modular fractal model, here we have the additional
requirement that the sum of all elements of the last eigenvector in the
Kronecker product has to
equal zero, i.e, $\sum_{i=0}^m v_i^k=0$ in Eq.~\eqref{eigvecyxc}. For details see Sec.~\ref{app2} in Appendix $A$, where we
derive Eq.~\eqref{eigvecyxc}.  This condition seems to be rather
strict. However, it is fulfilled for all the eigenvectors of all
motifs discussed in \cite{POE15} which are all generic, normalized motifs
of up to five nodes.  Once the eigensolutions are established their
stability can be studied as in \cite{POE15}. In special cases, if all
the eigenvectors in the Kronecker product are the same, 
hierarchical dynamics is obtained.


 

As an example, we investigate the initial motif $A_1$ described by
Eq. \eqref{adjm} for a hierarchy $n=3$. The number of nodes is given
by $5^{n}= 125$.  We consider the eigensolution given by $v \otimes
v \otimes v$, where $v =  (1,1,0,-1-1)$ is an eigenvector of f
$A_1$. The motif is shown in Fig.~\ref{plko}(a), where the color scheme
indicates the state corresponding to $v = (1,1,0,-1-1)$. We observe that
the dynamics is identical to the dynamics shown in Fig.~\ref{plko}(b))
which we have obtained for the modular fractal model. This is due to the fact that in spite of very different topologies, the eigenvectors of the networks, and thus the dynamics, are identical.

We also study solutions that do not correspond to eigensolutions. We
do so for the three different motifs as shown in Fig.
\ref{motif_model1}(a), (b), and (c), respectively, and for a hierarchy
of $n=2$. The space-time plots are shown in Fig.
\ref{motif_model1}(d), (e), and (f), respectively.
 We observe that these solutions have as well a hierarchy
in their dynamics. The nodes of the middle motif and the middle node
of the remaining motifs are either phase shifted or have  different
radii than the remaining nodes or both. This is a result of the
structure of the motifs: The middle nodes of all three motifs have a
different connectivity than the other nodes. Additionally, in
Fig.~\ref{motif_model1} (f), the first two motifs have a different
amplitude than the last two, which is reflected in the horizontal
asymmetry in the topology about the central node. Thus, the
correlation between network topology and dynamics is not limited to
eigensolutions. 

\begin{figure}
  \centering
  \centering
  \noindent\stackinset{l}{}{t}{-0.6cm}{\colorbox{white}{(a)}}{
    \includegraphics[width=3cm]{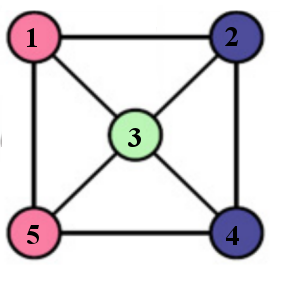} }
  \hfill
  \centering
  \noindent\stackinset{l}{}{t}{-0.6cm}{\colorbox{white}{(b)}}{
    \includegraphics[width=3cm]{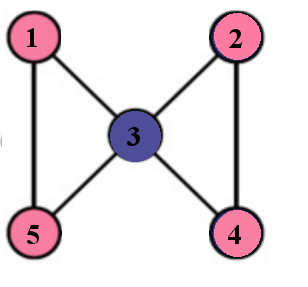} }
  \hfill
  \centering
  \noindent\stackinset{l}{}{t}{-0.6cm}{\colorbox{white}{(c)}}{
    \includegraphics[width=3cm]{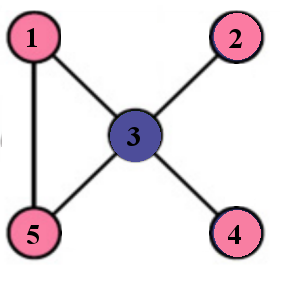} }
\noindent\stackinset{l}{}{t}{-0.6cm}{\colorbox{white}{(d)}}{
    \includegraphics[width=4cm, height=6cm]{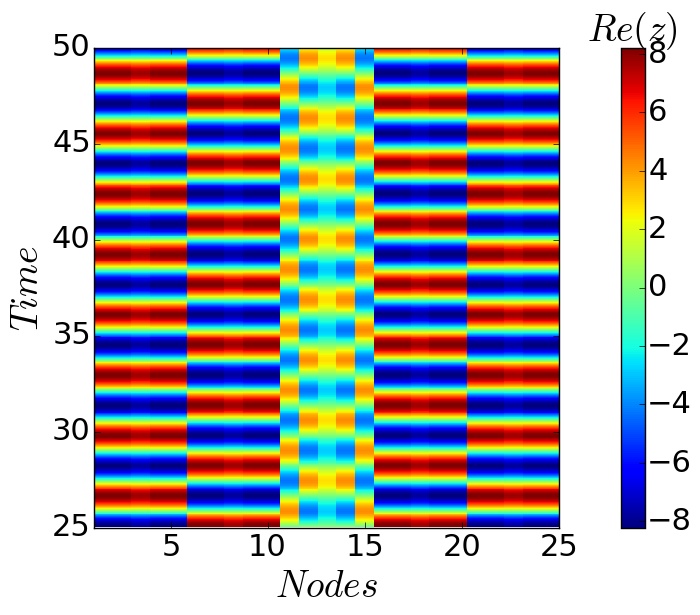} }
  \hfill
  \centering
  \noindent\stackinset{l}{}{t}{-0.6cm}{\colorbox{white}{(e)}}{
    \includegraphics[width=4cm, height=6cm]{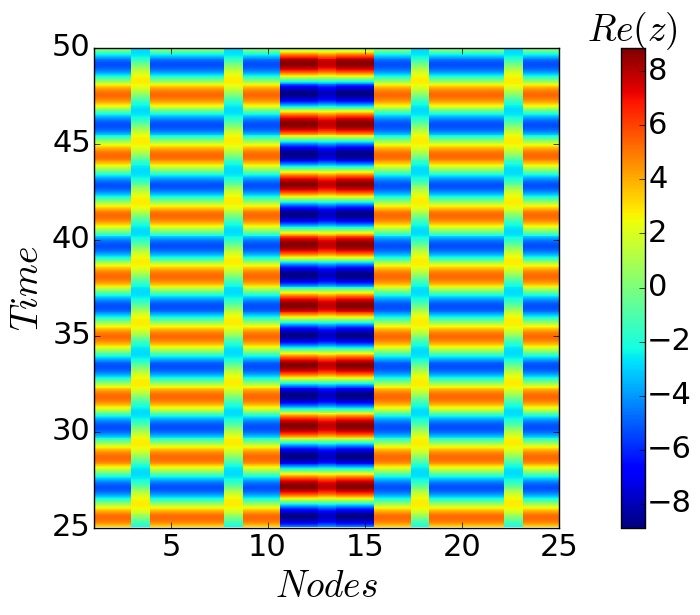} }
  \hfill
  \centering
  \noindent\stackinset{l}{}{t}{-0.6cm}{\colorbox{white}{(f)}}{
    \includegraphics[width=4cm, height=6cm]{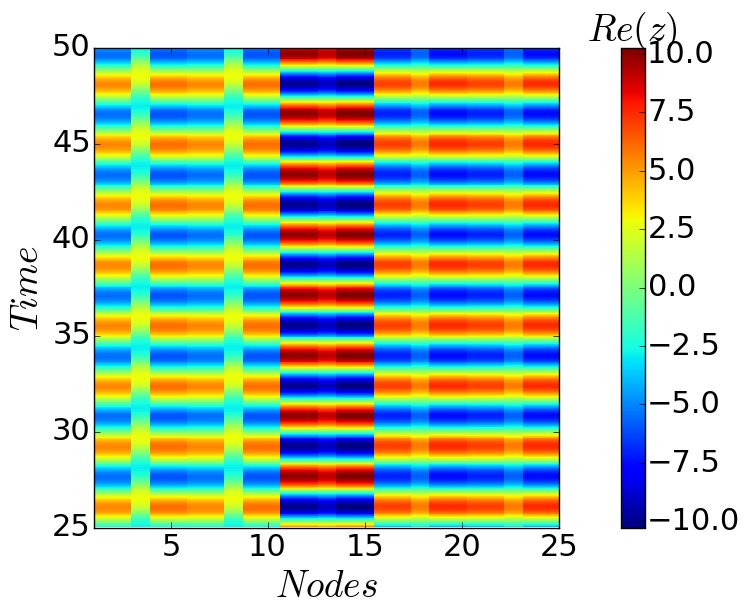} }
  \vfill
    \caption{ Non-eigensolution dynamics in the hierarchical model.
         (a), (b) and (c): initial motifs. Colors denote the eigensolutions of the motif
    (which do not directly translate to the larger network).
    The green node shows amplitude death, the
    red and blue nodes are antiphase-synchronized with respect to each
    other.
(d), (e), and (f) are space-time plots for the networks built from the
initial motifs in (a), (b), and (c),
    respectively.
    Other parameters: $n = 2$, $N = 5^2 = 25$, $ \sigma = -5, \beta=0, \lambda=1, \omega=2$.}
  \label{motif_model1}
\end{figure}

\section{Conclusion}
\label{concl}
In this paper we have presented three different models of networks with hierarchical or fractal connectivities, and studied them analytically using an eigensolution
concept. The eigensolution concept was developed in \cite{POE15} to
describe synchronization, anti-synchronization, and partial amplitude
death. Here, we extend this concept to cluster states and to larger
hierarchical networks created from basic motifs. In combination
with hierarchical topologies this leads to complex
synchronization patterns such as cluster synchronization, partial
synchronization, oscillation death, partial amplitude death and nested dynamics. In particular, we observe that the fractal nature of
the network translates to fractal synchronization
patterns. Understanding these synchronization patterns helps to
bridge the gap between relatively simple, by now well understood states like
in-phase synchronization and much more complex synchronization patterns
like chimera states.

The first model we have considered has a one-dimensional fractal
topology with a circulant adjacency matrix, and the eigensolutions are
cluster states. For networks with circulant adjacency matrix, we have
mathematically derived the conditions for the existence of partial
amplitude death eigensolutions and calculated their stability.  We
have also shown that oscillation death, possibly coexisting with partial amplitude death, arises in such networks as a special
cluster state with zero-frequency.  This is in contrast to previous
work, where oscillation death was observed as a result of symmetry
breaking. Therefore, we establish here a second mechanism leading to oscillation death.

The second model has a two-dimensional modular fractal topology,
and the third model is a direct extension of mesoscale motifs to
larger self-similar hierarchical networks. In both the second and the third model we see the direct
influence of topology on dynamics, i.e., for a hierarchical topology,
we obtain hierarchical dynamics. Although the second and the third model yield similar dynamics, they have vastly different topologies that resemble very different natural systems and hence, have different applications. The second model has a fractal hierarchical topology whereas the third model has a self-similar hierarchical topology. In addition, while the second model is applicable to networks created from the mesoscale motifs in \cite{POE15} as well as to all networks with circulant adjacency matrices, the third model is relevant only for the former.

The work presented here is of particular interest for neuroscience
where recently a lot of emphasis has been put on the relation between
structural connectivity and functional connectivity in the
brain\cite{RUB09, STA15, ABD14, COM11}. Evidence from empirical studies suggests that the presence of a direct anatomical connection between two brain areas is associated with stronger functional interactions between these two areas \cite{HON07, HON09, RUB09, HER13}. Our results support these empirical results through theoretical investigation. In addition, they can give
valuable insight because they provide a completely analytical
framework while employing a complex hierarchical
structure that mimics the hierarchical nature of neurons in the brain
\cite{OME15, KAT09, EXP11, KAT12}. The fractal or self-similar
hierarchical organization of neural networks is studied in \cite{KAI07, ROB09, SPO06, DOU04}. The advantage of this theoretical study is that it allows for investigating the
interplay of dynamics and topology on every scale, from the smallest
to the largest structural level as well as the investigation of dynamics of each individual node. It is therefore a powerful complement
to experimental work, which, due to its challenging nature, is often
limited to mean-field approximations\cite{BOJ10}, and to theoretical work representing
neural dynamics in terms of overall statistics, i.e., representing
entire cortical regions as one node \cite{ZHO06,BOJ10,PIN14,SAN15a}. Besides
applying it to neuroscience, our work can also be used
to study the functional dynamics of metabolic networks, which also
have been shown to display a hierarchical topology as in our third model \cite{RAV02}.

This correlation between dynamics and topology is not limited to the
eigensolutions. We have shown that synchronization patterns corresponding to more general solutions can be predicted in hierarchical networks from the knowledge of motif topology. The study of these solutions and the extension of the
eigensolution concept to time-delayed networks would be an interesting topic of further research.

\aucontribute{S. Krishnagopal introduced the models and performed the analysis and simulations under the supervision of J. Lehnert and E. Schöll. The analytical results in this paper are based on the eigensolution technique developed by W. Poel under the supervision of A. Zakharova  and E. Schöll \cite{POE15}. All authors contributed to writing the manuscript.}

\competing{'The author(s) declare that they have no competing interests’.}

\funding{This work was supported by DFG in the framework of SFB 910: Control of self-organizing nonlinear systems: Theoretical methods and concepts of application}


\section{Appendix A}

The proofs from Sec.~\ref{results} are given here.

\subsection{A.1 : Eigenvectors of networks with modular fractal topology}
\label{app1}

Here we will discuss how the eigenvectors of the second model can be
calculated from the initial motif, i.e., we derive Eq.~\eqref{eigvecasd}.
Let us consider the initial matrix $A_1$ of size $m \times m$. The final adjacency
matrix for the hierarchy level $n=2$ is given by $A = A_1\otimes
A_1$. Let $v^i$, and $v^j$ be two of the eigenvectors of $A_1$. Then,
$A_1v^i = \lambda_i v^i$ and $A_1 v^j = \lambda_j v^j$ holds. In order for $v^i \otimes v^j, \quad \text{where} \quad i,j \in \{1, \ldots m \}$, to be an eigenvector of the above matrix, the following equation has to be satisfied:
\begin{equation}
(A_1 \otimes A_1 - \lambda) (v^i \otimes v^j)= 0.
\label{model2}
\end{equation}
We make use of the following  property of the Kronecker product:
$(H \otimes I)(J \otimes K) = HJ \otimes IK $, where $H$, $I$, $J$,
and $K$ are arbitrary matrices of appropriate dimensions. Using in
addition $A_1v^k=\lambda_k v^k, k \in \{i,j\}$, Eq. \eqref{model2} evaluates to: 
\begin{align}
\lambda_i \lambda_j v^i \otimes v^j - \lambda v^i \otimes v^j = 0, \quad \quad \text{for } \lambda=\lambda_j  \lambda_i.
\label{asd}
\end{align}

Thus, $v^i \otimes v^j$ is an eigenvector of $A$. The corresponding eigenvalue is given by $ \lambda = \lambda_j \lambda_i $.

The adjacency matrix for the $n^{\text{th}}$ hierarchy is obtained by
repeating the procedure  above $n-1$ times and replacing $A_1$ with the
matrix of the previous iteration each time. We,
thus, obtain
\begin{equation}
v=\overbrace{v^i\otimes v^j\otimes \ldots \otimes v^k, }^\text{$n$ times} \quad \text{where} \quad i,j,k \in \{1, \ldots m \},
\end{equation} 
which is equivalent to Eq.~\eqref{eigvecasd}.
\subsection{A.2 : Eigenvectors of networks with hierarchical topology}
\label{app2}
Here, we derive the relation for the eigenvector of the third model, i.e., Eq.~\eqref{eigvecyxc}.
Let us consider an initial matrix $A_1$ of size $m \times m$. The final adjacency matrix for the hierarchy level $n=2$ is given by $A_{m^2} = A_1 \otimes E_m + \mathbf{1_m} \otimes A_1$ where $E_m$ is a matrix of size $m \times m$ where all the entries are 1, and $\mathbf{1_m}$ is the $m \times m$ identity matrix.
$v^i$, and $v^j$ are eigenvectors of $A_1$: $A_1v^i = \lambda_i v^i,\quad A_1 v^j = \lambda_i v^j.$ In order for $v^i \otimes v^j, \quad \text{where} \quad i,j \in \{1, \ldots m \}$, to be an eigenvector of the above matrix, the following equation has to be satisfied:
\begin{equation}
(A_1 \otimes E_m + \mathbf{1_m} \otimes A_1 - \lambda) (v^i \otimes v^j)= 0.
\label{zoo}
\end{equation}
 \newline
Using the property of the Kronecker product
$(H \otimes I)(J \otimes K) = HJ \otimes IK $, where $H$, $I$, $J$,
and $K$ are arbitrary matrices of appropriate dimensions, and $A_1v^k=\lambda_k v^k, k \in \{i,j\}$, Eq.~\eqref{zoo} evaluates to
\begin{align}
A_1 v^i \otimes \left(\sum_{i=1}^mv_i^j\right) o_m + v^i \otimes A_1 v^j - \lambda v^i \otimes v^j = 0,
\end{align}
where $o_m$ is a vector of ones of size $m$. If $\sum_{i=1}^mv_i^j = 0$, this reduces to
\begin{align} 
\lambda_j v^i \otimes v^j - \lambda v^i \otimes v^j
= 0 \quad \text{for} \quad \lambda=\lambda_j.
\end{align}

Thus, $v^i \otimes v^j$ is an eigenvector of A. The corresponding
eigenvalue is given by $ \lambda = \lambda_j$. The adjacency matrix
$A$ for the $n^{\text{th}}$ level of the hierarchy  is obtained by repeating the
above $n-1$ times and replacing $A_1$ with the
matrix of the previous iteration each time. Doing so we obtain Eq.~\eqref{eigvecyxc}. 
\subsection{A.3: Example of a modular fractal adjacency matrix}
For the five-node motif shown in Fig.~\ref{plo}(a) \cite{POE15} and described by the adjacency matrix $A_1$ (Eq.~\eqref{adjm}),
the final adjacency matrix $A$ for $n=2$, i.e., $ A = A_1 \otimes A_1$ is given explicitly by

\begin{equation}
A_1 = \left(
\begin{array}{@{}*{16}c@{}}
0 & 0 & 0 & 0 & 0 & 0           & \frac{1}{9} & \frac{1}{9} & 0 & \frac{1}{9} & \ldots & 0           & \frac{1}{9} & \frac{1}{9} & 0 & \frac{1}{9}\\
0 & 0 & 0 & 0 & 0 & \frac{1}{9} &0& \frac{1}{9}            &
\frac{1}{9} & 0 & \ldots& \frac{1}{9} &0& \frac{1}{9}           &
\frac{1}{9} & 0   \\
0 & 0 & 0 & 0 & 0 &  \frac{1}{12} &\frac{1}{12} &0 & \frac{1}{12} & \frac{1}{12}&\ldots &    \frac{1}{12} &\frac{1}{12} &0 & \frac{1}{12}& \frac{1}{12}\\
0 & 0 & 0 & 0 & 0 & 0 &    \frac{1}{9} &\frac{1}{9} &0 & \frac{1}{9} &\ldots &0 &    \frac{1}{9} &\frac{1}{9} &0 & \frac{1}{9}\\
0 & 0 & 0 & 0 & 0 &     \frac{1}{9}&0 &\frac{1}{9} & \frac{1}{9} &0 &\ldots &  \frac{1}{9}&0 &\frac{1}{9} & \frac{1}{9}&0\\
 0           & \frac{1}{9} & \frac{1}{9} & 0 &
\frac{1}{9} & 0 & 0 & 0 & 0 & 0 &\ldots & 0 & 0 & 0 & 0 & 0 \\
 \frac{1}{9} &0& \frac{1}{9}           &
\frac{1}{9} & 0 &0 & 0 & 0 & 0 & 0 & \ldots& 0 & 0 & 0 & 0 & 0   \\
\frac{1}{12} &\frac{1}{12} &0 & \frac{1}{12} & \frac{1}{12}&0 & 0 & 0 & 0 & 0 &\ldots &    0 & 0 & 0 & 0 & 0 \\
0 &    \frac{1}{9} &\frac{1}{9} &0 & \frac{1}{9}  & 0 & 0 & 0 & 0 &0&\ldots &0 &0 &0 &0 & 0\\
 \frac{1}{9}&0 &\frac{1}{9} & \frac{1}{9} &0 &0 & 0 & 0 & 0 & 0 &\ldots & 0 & 0 & 0 & 0 & 0 \\
\vdots & \vdots & \vdots & \vdots & \vdots & \vdots           & \vdots & \vdots & \vdots &
\vdots & \ldots & \vdots & \vdots & \vdots & \vdots & \vdots \\
 0           & \frac{1}{9} & \frac{1}{9} & 0 &
\frac{1}{9} & 0 & 0 & 0 & 0 & 0 &\ldots & 0 & 0 & 0 & 0 & 0 \\
 \frac{1}{9} &0& \frac{1}{9}           &
\frac{1}{9} & 0 &0 & 0 & 0 & 0 & 0 & \ldots& 0 & 0 & 0 & 0 & 0   \\
\frac{1}{12} &\frac{1}{12} &0 & \frac{1}{12} & \frac{1}{12}&0 & 0 & 0 & 0 & 0 &\ldots &    0 & 0 & 0 & 0 & 0 \\
0 &    \frac{1}{9} &\frac{1}{9} &0 & \frac{1}{9}  & 0 & 0 & 0 & 0 &0&\ldots &0 &0 &0 &0 & 0\\
 \frac{1}{9}&0 &\frac{1}{9} & \frac{1}{9} &0 &0 & 0 & 0 & 0 & 0 &\ldots & 0 & 0 & 0 & 0 & 0 \\
\end{array}\right).
\label{adjmA}
\end{equation}

%

\end{document}

%% file: diss_cluster.pdf_t
\begin{picture}(0,0)%
\includegraphics{diss_cluster.pdf}%
\end{picture}%
\setlength{\unitlength}{2881sp}%
\begingroup\makeatletter\ifx\SetFigFont\undefined%
\gdef\SetFigFont#1#2#3#4#5{%
  \reset@font\fontsize{#1}{#2pt}%
  \fontfamily{#3}\fontseries{#4}\fontshape{#5}%
  \selectfont}%
\fi\endgroup%
\begin{picture}(8955,2072)(54,-16124)
\put(7332,-14549){\makebox(0,0)[lb]{\smash{{\SetFigFont{11}{13.2}{\rmdefault}{\mddefault}{\updefault}{\color[rgb]{0.000,0.200,0.400}$\frac{3\pi} {2}$}%
}}}}
\put(5717,-14259){\makebox(0,0)[lb]{\smash{{\SetFigFont{11}{13.2}{\rmdefault}{\mddefault}{\updefault}{\color[rgb]{0.000,0.200,0.400}$\pi$}%
}}}}
\put(4009,-14565){\makebox(0,0)[lb]{\smash{{\SetFigFont{11}{13.2}{\rmdefault}{\mddefault}{\updefault}{\color[rgb]{0.000,0.200,0.400}$\frac{\pi} {2}$}%
}}}}
\put(3369,-15962){\makebox(0,0)[lb]{\smash{{\SetFigFont{11}{13.2}{\rmdefault}{\mddefault}{\updefault}{\color[rgb]{0,0,0}$z_4$}%
}}}}
\put(8994,-15137){\makebox(0,0)[lb]{\smash{{\SetFigFont{11}{13.2}{\rmdefault}{\mddefault}{\updefault}{\color[rgb]{0,0,0}$z_{1}$}%
}}}}
\put(8718,-15737){\makebox(0,0)[lb]{\smash{{\SetFigFont{11}{13.2}{\rmdefault}{\mddefault}{\updefault}{\color[rgb]{0,0,0}$\tilde{\omega}$}%
}}}}
\put( 69,-14387){\makebox(0,0)[lb]{\smash{{\SetFigFont{11}{13.2}{\rmdefault}{\mddefault}{\updefault}{\color[rgb]{0,0,0}$(a$)}%
}}}}
\put(4194,-15137){\makebox(0,0)[lb]{\smash{{\SetFigFont{11}{13.2}{\rmdefault}{\mddefault}{\updefault}{\color[rgb]{0,0,0}$z_{1}$}%
}}}}
\put(3369,-14312){\makebox(0,0)[lb]{\smash{{\SetFigFont{11}{13.2}{\rmdefault}{\mddefault}{\updefault}{\color[rgb]{0,0,0}$z_2$}%
}}}}
\put(2169,-14387){\makebox(0,0)[lb]{\smash{{\SetFigFont{11}{13.2}{\rmdefault}{\mddefault}{\updefault}{\color[rgb]{0,0,0}$(b)$}%
}}}}
\put(3069,-14837){\makebox(0,0)[lb]{\smash{{\SetFigFont{11}{13.2}{\rmdefault}{\mddefault}{\updefault}{\color[rgb]{0,0,0}$r_o$}%
}}}}
\put(4569,-14387){\makebox(0,0)[lb]{\smash{{\SetFigFont{11}{13.2}{\rmdefault}{\mddefault}{\updefault}{\color[rgb]{0,0,0}$(c)$}%
}}}}
\put(6969,-14387){\makebox(0,0)[lb]{\smash{{\SetFigFont{11}{13.2}{\rmdefault}{\mddefault}{\updefault}{\color[rgb]{0,0,0}$(d)$}%
}}}}
\put(7869,-14837){\makebox(0,0)[lb]{\smash{{\SetFigFont{11}{13.2}{\rmdefault}{\mddefault}{\updefault}{\color[rgb]{0,0,0}$r_o$}%
}}}}
\put(5844,-14987){\makebox(0,0)[lb]{\smash{{\SetFigFont{11}{13.2}{\rmdefault}{\mddefault}{\updefault}{\color[rgb]{0,0,0}$r_o$}%
}}}}
\put(3909,-15737){\makebox(0,0)[lb]{\smash{{\SetFigFont{11}{13.2}{\rmdefault}{\mddefault}{\updefault}{\color[rgb]{0,0,0}$\tilde{\omega}$}%
}}}}
\put(1509,-15752){\makebox(0,0)[lb]{\smash{{\SetFigFont{11}{13.2}{\rmdefault}{\mddefault}{\updefault}{\color[rgb]{0,0,0}$\tilde{\omega}$}%
}}}}
\put(6309,-15752){\makebox(0,0)[lb]{\smash{{\SetFigFont{11}{13.2}{\rmdefault}{\mddefault}{\updefault}{\color[rgb]{0,0,0}$\tilde{\omega}$}%
}}}}
\put(2469,-15137){\makebox(0,0)[lb]{\smash{{\SetFigFont{11}{13.2}{\rmdefault}{\mddefault}{\updefault}{\color[rgb]{0,0,0}$z_{3}$}%
}}}}
\put(4601,-15137){\makebox(0,0)[lb]{\smash{{\SetFigFont{11}{13.2}{\rmdefault}{\mddefault}{\updefault}{\color[rgb]{0,0,0}$z_2,z_4$}%
}}}}
\put(519,-16037){\makebox(0,0)[lb]{\smash{{\SetFigFont{11}{13.2}{\rmdefault}{\mddefault}{\updefault}{\color[rgb]{0,0,0}$j=1, \dots, 4 $}%
}}}}
\put(8169,-15962){\makebox(0,0)[lb]{\smash{{\SetFigFont{11}{13.2}{\rmdefault}{\mddefault}{\updefault}{\color[rgb]{0,0,0}$z_2$}%
}}}}
\put(7290,-15134){\makebox(0,0)[lb]{\smash{{\SetFigFont{11}{13.2}{\rmdefault}{\mddefault}{\updefault}{\color[rgb]{0,0,0}$z_{3}$}%
}}}}
\put(8169,-14312){\makebox(0,0)[lb]{\smash{{\SetFigFont{11}{13.2}{\rmdefault}{\mddefault}{\updefault}{\color[rgb]{0,0,0}$z_4$}%
}}}}
\put(1053,-15007){\makebox(0,0)[lb]{\smash{{\SetFigFont{11}{13.2}{\rmdefault}{\mddefault}{\updefault}{\color[rgb]{0,0,0}$r_o$}%
}}}}
\put(1775,-15129){\makebox(0,0)[lb]{\smash{{\SetFigFont{11}{13.2}{\rmdefault}{\mddefault}{\updefault}{\color[rgb]{0,0,0}$z_j$}%
}}}}
\put(6594,-15137){\makebox(0,0)[lb]{\smash{{\SetFigFont{11}{13.2}{\rmdefault}{\mddefault}{\updefault}{\color[rgb]{0,0,0}$z_1,z_3$}%
}}}}
\end{picture}%

%% file: diss_cluster_2.pdf_t
\begin{picture}(0,0)%
\includegraphics{diss_cluster_2.pdf}%
\end{picture}%
\setlength{\unitlength}{2881sp}%
\begingroup\makeatletter\ifx\SetFigFont\undefined%
\gdef\SetFigFont#1#2#3#4#5{%
  \reset@font\fontsize{#1}{#2pt}%
  \fontfamily{#3}\fontseries{#4}\fontshape{#5}%
  \selectfont}%
\fi\endgroup%
\begin{picture}(3105,1927)(5986,-16118)
\put(7201,-14611){\makebox(0,0)[lb]{\smash{{\SetFigFont{11}{13.2}{\rmdefault}{\mddefault}{\updefault}{\color[rgb]{0.000,0.200,0.400}$\frac{3\pi} {2}$}%
}}}}
\put(8002,-14966){\makebox(0,0)[lb]{\smash{{\SetFigFont{11}{13.2}{\rmdefault}{\mddefault}{\updefault}{\color[rgb]{0,0,0}$r_o$}%
}}}}
\put(9076,-15361){\makebox(0,0)[lb]{\smash{{\SetFigFont{11}{13.2}{\rmdefault}{\mddefault}{\updefault}{\color[rgb]{0,0,0}$z_{1}, z_{5},z_{9},z_{13}$}%
}}}}
\put(6001,-15061){\makebox(0,0)[lb]{\smash{{\SetFigFont{11}{13.2}{\rmdefault}{\mddefault}{\updefault}{\color[rgb]{0,0,0}$z_{3}, z_{7},z_{11},z_{15}$}%
}}}}
\put(7876,-14386){\makebox(0,0)[lb]{\smash{{\SetFigFont{11}{13.2}{\rmdefault}{\mddefault}{\updefault}{\color[rgb]{0,0,0}$z_{2}, z_{6},z_{10},z_{14}$}%
}}}}
\put(6751,-16036){\makebox(0,0)[lb]{\smash{{\SetFigFont{11}{13.2}{\rmdefault}{\mddefault}{\updefault}{\color[rgb]{0,0,0}$z_{4}, z_{8},z_{12},z_{16}$}%
}}}}
\put(8626,-15961){\makebox(0,0)[lb]{\smash{{\SetFigFont{11}{13.2}{\rmdefault}{\mddefault}{\updefault}{\color[rgb]{0,0,0}$\tilde{\omega} \neq 0$}%
}}}}
\end{picture}%

%% file: diss_cluster_1.pdf_t
\begin{picture}(0,0)%
\includegraphics{diss_cluster_1.pdf}%
\end{picture}%
\setlength{\unitlength}{2881sp}%
\begingroup\makeatletter\ifx\SetFigFont\undefined%
\gdef\SetFigFont#1#2#3#4#5{%
  \reset@font\fontsize{#1}{#2pt}%
  \fontfamily{#3}\fontseries{#4}\fontshape{#5}%
  \selectfont}%
\fi\endgroup%
\begin{picture}(2955,2077)(6136,-16118)
\put(7332,-14549){\makebox(0,0)[lb]{\smash{{\SetFigFont{11}{13.2}{\rmdefault}{\mddefault}{\updefault}{\color[rgb]{0.000,0.200,0.400}$\frac{3\pi} {2}$}%
}}}}
\put(6151,-15136){\makebox(0,0)[lb]{\smash{{\SetFigFont{11}{13.2}{\rmdefault}{\mddefault}{\updefault}{\color[rgb]{0,0,0}$z_{3}, z_{7},z_{11},z_{15}$}%
}}}}
\put(8776,-15736){\makebox(0,0)[lb]{\smash{{\SetFigFont{11}{13.2}{\rmdefault}{\mddefault}{\updefault}{\color[rgb]{0,0,0}$\tilde{\omega}=0$}%
}}}}
\put(8026,-14836){\makebox(0,0)[lb]{\smash{{\SetFigFont{11}{13.2}{\rmdefault}{\mddefault}{\updefault}{\color[rgb]{0,0,0}$r_o$}%
}}}}
\put(7426,-14236){\makebox(0,0)[lb]{\smash{{\SetFigFont{11}{13.2}{\rmdefault}{\mddefault}{\updefault}{\color[rgb]{0,0,0}$z_{2}, z_{6},z_{10},z_{14}$}%
}}}}
\put(7426,-16036){\makebox(0,0)[lb]{\smash{{\SetFigFont{11}{13.2}{\rmdefault}{\mddefault}{\updefault}{\color[rgb]{0,0,0}$z_{4}, z_{8},z_{12},z_{16}$}%
}}}}
\put(9076,-15136){\makebox(0,0)[lb]{\smash{{\SetFigFont{11}{13.2}{\rmdefault}{\mddefault}{\updefault}{\color[rgb]{0,0,0}$z_{1}, z_{5},z_{9},z_{13}$}%
}}}}
\end{picture}%

%% file: kri16.bbl
\begin{thebibliography}{10}
\expandafter\ifx\csname url\endcsname\relax
  \def\url#1{{\tt #1}}\fi
\expandafter\ifx\csname urlprefix\endcsname\relax\def\urlprefix{URL }\fi

\bibitem{SCH16}
E.~Sch{\"o}ll, S.~H.~L. Klapp, and P.~H{\"o}vel: {\em Control of
  self-organizing nonlinear systems\/} (Springer, Berlin, 2016).

\bibitem{GAR04}
J.~Garc{\'i}a-Ojalvo, M.~B. Elowitz, and S.~H. Strogatz: {\em Modeling a
  synthetic multicellular clock: Repressilators coupled by quorum sensing\/},
  PNAS {\bf 101}, 10955 (2004).

\bibitem{BLA99a}
B.~Blasius, A.~Huppert, and L.~Stone: {\em Complex dynamics and phase
  synchronization in spatially extended ecological systems\/}, Nature (London)
  {\bf 399}, 354 (1999).

\bibitem{MIY07}
T.~Miyano and T.~Tsutsui: {\em Data synchronization in a network of coupled
  phase oscillators\/}, Phys. Rev. Lett. {\bf 98}, 024102 (2007).

\bibitem{WIT12}
D.~Witthaut and M.~Timme: {\em Braess's paradox in oscillator networks,
  desynchronization and power outage\/}, New Journal of Physics {\bf 14},
  083036 (2012).

\bibitem{ROH12}
M.~Rohden, A.~Sorge, M.~Timme, and D.~Witthaut: {\em Self-organized
  synchronization in decentralized power grids\/}, Phys. Rev. Lett. {\bf 109},
  064101 (2012).

\bibitem{PLU05}
A.~Pluchino, V.~Latora, and A.~Rapisarda: {\em Changing opinions in a changing
  world: A new perspective in sociophysics\/}, International Journal of Modern
  Physics C {\bf 16}, 515 (2005), cited By 29.

\bibitem{GOL02a}
M.~Golubitsky and I.~Stewart: {\em {The symmetry perspective}\/}
  (Birkh{\"a}user, Basel, 2002).

\bibitem{SOR07}
F.~Sorrentino and E.~Ott: {\em Network synchronization of groups\/}, Phys.
  Rev.~E {\bf 76}, 056114 (2007).

\bibitem{KES08}
J.~Kestler, E.~Kopelowitz, I.~Kanter, and W.~Kinzel: {\em Patterns of chaos
  synchronization\/}, Phys.~Rev.~E {\bf 77}, 046209 (2008).

\bibitem{KAN11}
I.~Kanter, M.~Zigzag, A.~Englert, F.~Geissler, and W.~Kinzel: {\em
  Synchronization of unidirectional time delay chaotic networks and the
  greatest common divisor\/}, Europhys.~Lett. {\bf 93}, 60003 (2011).

\bibitem{ILL11}
L.~Illing, C.~D. Panda, and L.~Shareshian: {\em Isochronal chaos
  synchronization of delay-coupled optoelectronic oscillators\/}, Phys. Rev. E
  {\bf 84}, 016213 (2011).

\bibitem{DAH12}
T.~Dahms, J.~Lehnert, and E.~Sch{\"o}ll: {\em Cluster and group synchronization
  in delay-coupled networks\/}, Phys. Rev.~E {\bf 86}, 016202 (2012).

\bibitem{LUE12a}
L.~L{\"u}cken and S.~Yanchuk: {\em Two-cluster bifurcations in systems of
  globally pulse-coupled oscillators\/}, Physica~D {\bf 241}, 350 (2012).

\bibitem{BLA13}
K.~Blaha, J.~Lehnert, A.~Keane, T.~Dahms, P.~H{\"o}vel, E.~Sch{\"o}ll, and
  J.~L. Hudson: {\em Clustering in delay-coupled smooth and relaxational
  chemical oscillators\/}, Phys. Rev.~E {\bf 88}, 062915 (2013).

\bibitem{WIL12a}
C.~R.~S. Williams, F.~Sorrentino, T.~E. Murphy, R.~Roy, T.~Dahms, and
  E.~Sch{\"o}ll: {\em Group synchrony in an experimental system of
  delay-coupled optoelectronic oscillators\/}, in {\em Proc. 2012 Internat.
  Symposium on Nonlinear Theory and its Applications (NOLTA2012), Palma de
  Mallorca\/} (IEICE, Japan, 2012), pp. 70--73.

\bibitem{WIL13}
C.~R.~S. Williams, T.~E. Murphy, R.~Roy, F.~Sorrentino, T.~Dahms, and
  E.~Sch{\"o}ll: {\em Experimental observations of group synchrony in a system
  of chaotic optoelectronic oscillators\/}, Phys. Rev. Lett. {\bf 110}, 064104
  (2013).

\bibitem{PEC14}
L.~M. Pecora, F.~Sorrentino, A.~M. Hagerstrom, T.~E. Murphy, and R.~Roy: {\em
  Symmetries, cluster synchronization, and isolated desynchronization in
  complex networks\/}, Nature Commun. {\bf 5}, 4079 (2014).

\bibitem{ROS15}
D.~P. Rosin: {\em Dynamics of Complex Autonomous Boolean Networks\/}, Springer
  Theses (Springer, Heidelberg, 2015).

\bibitem{SOR16a}
F.~Sorrentino, L.~M. Pecora, A.~M. Hagerstrom, T.~E. Murphy, and R.~Roy: {\em
  Complete characterization of the stability of cluster synchronization in
  complex dynamical networks\/}, Sci. Adv. {\bf 2}, e1501737 (2016).

\bibitem{MOH06}
P.~K. Mohanty and A.~Politi: {\em {A new approach to partial synchronization in
  globally coupled rotators}\/}, Journal of Physics A: Mathematical and General
  {\bf 39}, L415 (2006).

\bibitem{WAG02a}
D.~J. {Wagg}: {\em {Partial Synchronization of Nonidentical Chaotic Systems via
  Adaptive Control, with Applications to Modeling Coupled Nonlinear
  Systems}\/}, International Journal of Bifurcation and Chaos {\bf 12}, 561
  (2002).

\bibitem{RUB02}
L.~L. Rubchinsky, M.~M. Sushchik, and G.~V. Osipov: {\em Patterns in networks
  of oscillators formed via synchronization and oscillator death\/}, Math.
  Comp. Simul {\bf 58}, 443 (2002).

\bibitem{POE15}
W.~Poel, A.~Zakharova, and E.~Sch{\"o}ll: {\em Partial synchronization and
  partial amplitude death in mesoscale network motifs\/}, Phys. Rev.~E {\bf
  91}, 022915 (2015).

\bibitem{ATA02a}
F.~M. Atay: {\em Total and partial amplitude death in networks of diffusively
  coupled oscillators\/}, Physica D {\bf 183}, 1 (2002).

\bibitem{SAX12}
G.~Saxena, A.~Prasad, and R.~Ramaswamy: {\em Amplitude death: The emergence of
  stationarity in coupled nonlinear systems\/}, Phys. Rep. {\bf 521}, 205
  (2012).

\bibitem{KOS13}
A.~Koseska, E.~Volkov, and J.~Kurths: {\em Oscillation quenching mechanisms:
  Amplitude vs. oscillation death\/}, Phys. Rep. {\bf 531}, 173 (2013).

\bibitem{KUR02a}
Y.~Kuramoto and D.~Battogtokh: {\em {Coexistence of Coherence and Incoherence
  in Nonlocally Coupled Phase Oscillators.}\/}, Nonlin. Phen. in Complex Sys.
  {\bf 5}, 380 (2002).

\bibitem{ABR04}
D.~M. Abrams and S.~H. Strogatz: {\em Chimera states for coupled
  oscillators\/}, Phys.~Rev.~Lett. {\bf 93}, 174102 (2004).

\bibitem{HAG12}
A.~M. Hagerstrom, T.~E. Murphy, R.~Roy, P.~H{\"o}vel, I.~Omelchenko, and
  E.~Sch{\"o}ll: {\em Experimental observation of chimeras in coupled-map
  lattices\/}, Nature Phys. {\bf 8}, 658 (2012).

\bibitem{TIN12}
M.~R. Tinsley, S.~Nkomo, and K.~Showalter: {\em Chimera and phase cluster
  states in populations of coupled chemical oscillators\/}, Nature Phys. {\bf
  8}, 662 (2012).

\bibitem{MAR13}
E.~A. Martens, S.~Thutupalli, A.~Fourriere, and O.~Hallatschek: {\em Chimera
  states in mechanical oscillator networks\/}, Proc. Nat. Acad. Sciences {\bf
  110}, 10563 (2013).

\bibitem{LAR13}
L.~Larger, B.~Penkovsky, and Y.~Maistrenko: {\em Virtual chimera states for
  delayed-feedback systems\/}, Phys. Rev. Lett. {\bf 111}, 054103 (2013).

\bibitem{OME13}
I.~Omelchenko, O.~Omel'chenko, P.~H{\"o}vel, and E.~Sch{\"o}ll: {\em When
  nonlocal coupling between oscillators becomes stronger: patched synchrony or
  multichimera states\/}, Phys. Rev. Lett. {\bf 110}, 224101 (2013).

\bibitem{ZAK14}
A.~Zakharova, M.~Kapeller, and E.~Sch{\"o}ll: {\em Chimera death: Symmetry
  breaking in dynamical networks\/}, Phys.~Rev.~Lett. {\bf 112}, 154101 (2014).

\bibitem{PAN15}
M.~J. Panaggio and D.~M. Abrams: {\em Chimera states: Coexistence of coherence
  and incoherence in networks of coupled oscillators\/}, Nonlinearity {\bf 28},
  R67 (2015).

\bibitem{OME16}
I.~Omelchenko, O.~Omel'chenko, A.~Zakharova, M.~Wolfrum, and E.~Sch{\"o}ll:
  {\em Tweezers for chimeras in small networks\/}, Phys. Rev. Lett. {\bf 116},
  114101 (2016).

\bibitem{SEM16}
N.~Semenova, A.~Zakharova, V.~S. Anishchenko, and E.~Sch{\"o}ll: {\em
  Coherence-resonance chimeras in a network of excitable elements\/}, Phys.
  Rev. Lett. {\bf 117}, 014102 (2016).

\bibitem{SCH16b}
E.~Sch{\"o}ll: {\em Synchronization patterns and chimera states in complex
  networks: interplay of topology and dynamics\/},
Eur. Phys. J. Spec. Top. {\bf 225}, 891
  (2016), {S}pecial Theme Issue on Mathematical Modeling of Complex Systems
  (ed. T. Bountis, A. Provata, G. Tsironis, J. Johnson).

\bibitem{KUM08}
P.~Kumar, A.~Prasad, and R.~Gosh: {\em Stable phase-locking of an
  external-cavity diode laser subjected to external optical injection\/}, J.
  Phys. B {\bf 41}, 135402 (2008).

\bibitem{ERM90}
G.~B. Ermentrout: {\em Oscillator death in populations of all to all coupled
  nonlinear oscillators\/}, Physica D {\bf 41}, 219 (1990).

\bibitem{CAK14}
C.~Cakan, J.~Lehnert, and E.~Sch{\"o}ll: {\em Heterogeneous delays in neural
  networks\/}, Eur. Phys.~J.~B {\bf 87}, 54 (2014).

\bibitem{SUZ11}
N.~Suzuki, C.~Furusawa, and K.~Kaneko: {\em Oscillatory protein expression
  dynamics endows stem cells with robust differentiation potential\/}, PLoS ONE
  {\bf 6}, e27232 (2011).

\bibitem{ZAK13a}
A.~Zakharova, I.~Schneider, Y.~N. Kyrychko, K.~B. Blyuss, A.~Koseska,
  B.~Fiedler, and E.~Sch{\"o}ll: {\em Time delay control of symmetry-breaking
  primary and secondary oscillation death\/}, Europhys. Lett. {\bf 104}, 50004
  (2013).

\bibitem{SCH15b}
I.~Schneider, M.~Kapeller, S.~Loos, A.~Zakharova, B.~Fiedler, and
  E.~Sch{\"o}ll: {\em Stable and transient multi-cluster oscillation death in
  nonlocally coupled networks\/}, Phys. Rev. E {\bf 92}, 052915 (2015).

\bibitem{PRO12}
A.~Provata, P.~Katsaloulis, and D.~A. Verganelakis: {\em Dynamics of chaotic
  maps for modelling the multifractal spectrum of human brain diffusion tensor
  images\/}, Chaos, Solitons \& Fractals {\bf 45}, 174 (2012).

\bibitem{OME15}
I.~Omelchenko, A.~Provata, J.~Hizanidis, E.~Sch{\"o}ll, and P.~H{\"o}vel: {\em
  Robustness of chimera states for coupled {FitzHugh-Nagumo} oscillators\/},
  Phys. Rev. E {\bf 91}, 022917 (2015).

\bibitem{HIZ15}
J.~Hizanidis, E.~Panagakou, I.~Omelchenko, E.~Sch{\"o}ll, P.~H{\"o}vel, and
  A.~Provata: {\em Chimera states in population dynamics: networks with
  fragmented and hierarchical connectivities\/}, Phys. Rev. E {\bf 92}, 012915
  (2015).

\bibitem{ULO16}
S.~Ulonska, I.~Omelchenko, A.~Zakharova, and E.~Sch{\"o}ll: {\em Chimera states
  in hierarchical networks of {Van der Pol} oscillators\/}, Chaos {\bf
  26}, 094825
  (2016).

\bibitem{TSI16}
N.~Tsigkri-DeSmedt, J.~Hizanidis, P.~H{\"o}vel, and A.~Provata: {\em
  Multi-chimera states and transitions in the leaky integrate-and-fire model
  with excitatory coupling and hierarchical connectivity\/}, Eur. Phys. J.
  Spec. Top.{\bf
  225}, 1149  (2016).

\bibitem{KAT09}
P.~Katsaloulis, D.~A. Verganelakis, and A.~Provata: {\em Fractal dimension and
  lacunarity of tractography images of the human brain\/}, Fractals {\bf 17},
  181 (2009).

\bibitem{EXP11}
P.~Expert, T.~S. Evans, V.~D. Blondel, and R.~Lambiotte: {\em Uncovering
  space-independent communities in spatial networks\/}, PNAS {\bf 108}, 7663
  (2011).

\bibitem{KAT12}
P.~Katsaloulis, A.~Ghosh, A.~C. Philippe, A.~Provata, and R.~Deriche: {\em
  Fractality in the neuron axonal topography of the human brain based on 3-d
  diffusion mri\/}, Eur. Phys. J.~B {\bf 85}, 1 (2012).

\bibitem{KAT12a}
P.~Katsaloulis, J.~Hizanidis, D.~A. Verganelakis, and A.~Provata: {\em
  Complexity measures and noise effects on {Diffusion Magnetic Resonance
  Imaging} of the neuron axons network in human brains\/}, Fluct. Noise Lett.
  {\bf 11}, 1250032 (2012).

\bibitem{SON05a}
C.~Song, S.~Havlin, and H.~A. Makse: {\em Self-similarity of complex
  networks\/}, Nature {\bf 433}, 392 (2005).

\bibitem{RAV02}
E.~Ravasz, D.~A. Mongru, Z.~N. Oltvai, and A.~L. Barabasi: {\em Hierarchical
  organization of modularity in metabolic networks\/}, Science {\bf 297}, 1551
  (2002).

\bibitem{DIL02}
S.~Dill, R.~Kumar, K.~S. Mccurley, S.~Rajagopalan, D.~Sivakumar, and
  A.~Tomkins: {\em Self-similarity in the web\/}, ACM Trans. Internet Technol.
  {\bf 2}, 205 (2002).

\bibitem{PEC98}
L.~M. Pecora and T.~L. Carroll: {\em Master stability functions for
  synchronized coupled systems\/}, Phys. Rev. Lett. {\bf 80}, 2109 (1998).

\bibitem{DHA04}
M.~Dhamala, V.~K. Jirsa, and M.~Ding: {\em Enhancement of neural synchrony by
  time delay\/}, Phys. Rev. Lett. {\bf 92}, 074104 (2004).

\bibitem{CHO09}
C.~U. Choe, T.~Dahms, P.~H{\"o}vel, and E.~Sch{\"o}ll: {\em Controlling
  synchrony by delay coupling in networks: from in-phase to splay and cluster
  states\/}, Phys. Rev.~E {\bf 81}, 025205(R) (2010).

\bibitem{FLU10b}
V.~Flunkert, S.~Yanchuk, T.~Dahms, and E.~Sch{\"o}ll: {\em Synchronizing
  distant nodes: a universal classification of networks\/}, Phys.~Rev.~Lett.
  {\bf 105}, 254101 (2010).

\bibitem{SOR13}
M.~C. Soriano, J.~Garc{\'i}a-Ojalvo, C.~R. Mirasso, and I.~Fischer: {\em
  Complex photonics: Dynamics and applications of delay-coupled semiconductors
  lasers\/}, Rev.~Mod.~Phys. {\bf 85}, 421 (2013).

\bibitem{WIL14}
C.~Wille, J.~Lehnert, and E.~Sch{\"o}ll: {\em Synchronization-desynchronization
  transitions in complex networks: An interplay of distributed time delay and
  inhibitory nodes\/}, Phys. Rev. E {\bf 90}, 032908 (2014).

\bibitem{TAY11}
A.~F. Taylor, M.~R. Tinsley, F.~Wang, and K.~Showalter: {\em Phase clusters in
  large populations of chemical oscillators\/}, Angew. Chem. {\bf 123}, 10343
  (2011), see also: Angew. Chem. Int. Ed. 50, 1-5 (2011).

\bibitem{DO12}
A.~L. Do, J.~M. H\"ofener, and T.~Gross: {\em Engineering mesoscale structures
  with distinct dynamical implications\/}, New J. of Phys. {\bf 14}, 115022
  (2012).

\bibitem{TAN14}
G.~Tanaka, K.~Morino, H.~Daido, and K.~Aihara: {\em Dynamical robustness of
  coupled heterogeneous oscillators\/}, Phys. Rev. E {\bf 89}, 052906 (2014).

\bibitem{MAT90}
P.~C. Matthews and S.~H. Strogatz: {\em Phase diagram for the collective
  behavior of limit-cycle oscillators\/}, Phys. Rev. Lett. {\bf 65}, 1701
  (1990).

\bibitem{HAK92}
V.~Hakim and W.~J. Rappel: {\em Dynamics of the globally coupled complex
  ginzburg-landau equation\/}, Phys. Rev. A {\bf 46}, R7347 (1992).

\bibitem{ATA03}
F.~M. Atay: {\em Distributed delays facilitate amplitude death of coupled
  oscillators\/}, Phys.~Rev.~Lett. {\bf 91}, 094101 (2003).

\bibitem{MIR90}
R.~E. Mirollo and S.~H. Strogatz: {\em Amplitude death in an array of
  limit-cycle oscillators\/}, J.~Stat.~Phys. {\bf 60}, 245 (1990).

\bibitem{OME11}
I.~Omelchenko, Y.~Maistrenko, P.~H{\"o}vel, and E.~Sch{\"o}ll: {\em Loss of
  coherence in dynamical networks: spatial chaos and chimera states\/}, Phys.
  Rev. Lett. {\bf 106}, 234102 (2011).

\bibitem{OME12}
I.~Omelchenko, B.~Riemenschneider, P.~H{\"o}vel, Y.~Maistrenko, and
  E.~Sch{\"o}ll: {\em Transition from spatial coherence to incoherence in
  coupled chaotic systems\/}, Phys. Rev.~E {\bf 85}, 026212 (2012).

\bibitem{DHU13}
O.~{D'Huys}, S.~Zeeb, T.~J{\"u}ngling, S.~Heiligenthal, S.~Yanchuk, and
  W.~Kinzel: {\em Synchronisation and scaling properties of chaotic networks
  with multiple delays\/}, EPL {\bf 103}, 10013 (2013).

\bibitem{GRA05}
R.~M. Gray: {\em Toeplitz and circulant matrices: A review.\/}, Found. Trends
  Commun. Inf. Theory {\bf 2} (2005).

\bibitem{TEE08}
G.~J. Tee: {\em Eigenvectors of block circulant and alternating circulant
  matrices\/}, Research Letters in the Information and Mathematical Sciences
  pp. 123--142 (2008).

\bibitem{RUB09}
M.~Rubinov, O.~Sporns, C.~Van~Leeuwen, and M.~Breakspear: {\em Symbiotic
  relationship between brain structure and dynamics\/}, BMC Neuroscience {\bf
  10}, 55 (2009).

\bibitem{STA15}
C.~J. Stam, E.~C.~W. van Straaten, E.~V. Dellen, P.~Tewarie, G.~Gong,
  A.~Hillebrand, J.~Meier, and P.~V. Mieghem: {\em {The relation between
  structural and functional connectivity patterns in complex brain
  networks}\/}, International Journal of Psychophysiology  (2015).

\bibitem{ABD14}
F.~Abdelnour, H.~U. Voss, and A.~Raj: {\em {Network diffusion accurately models
  the relationship between structural and functional brain connectivity
  networks}\/}, NeuroImage {\bf 90}, 335 (2014).

\bibitem{COM11}
P.~E.~C. Compeau, P.~A. Pevzner, and G.~Tesler: {\em How to apply de bruijn
  graphs to genome assembly\/}, Nature biotechnology {\bf 29}, 987 (2011).

\bibitem{HON07}
C.~J. Honey, R.~K{\"o}tter, M.~Breakspear, and O.~Sporns: {\em {{N}etwork
  structure of cerebral cortex shapes functional connectivity on multiple time
  scales}\/}, Proc. Natl. Acad. Sci. U.S.A. {\bf 104}, 10240 (2007).

\bibitem{HON09}
C.~J. Honey, O.~Sporns, L.~Cammoun, X.~Gigandet, J.~P. Thiran, R.~Meuli, and
  P.~Hagmann: {\em Predicting human resting-state functional connectivity from
  structural connectivity\/}, Proc. Natl. Acad. Sci. U.S.A. {\bf 106}, 2035
  (2009).

\bibitem{HER13}
A.~M. Hermundstad, D.~S. Bassett, K.~S. Brown, E.~M. Aminoff, and D.~Clewett:
  {\em Structural foundations of resting-state and task-based functional
  connectivity in the human brain\/}, Proceedings of the National Academy of
  Sciences {\bf 110}, 6169 (2013).

\bibitem{KAI07}
M.~Kaiser: {\em {Brain architecture: a design for natural computation}\/},
  Phil. Trans.~R. Soc.~A {\bf 365}, 3033 (2007).

\bibitem{ROB09}
P.~A. Robinson, J.~A. Henderson, E.~Matar, P.~Riley, and R.~T. Gray: {\em
  Dynamical reconnection and stability constraints on cortical network
  architecture\/}, Phys. Rev. Lett. {\bf 103}, 108104 (2009).

\bibitem{SPO06}
O.~Sporns: {\em {Small-world connectivity, motif composition, and complexity of
  fractal neuronal connections}\/}, Biosystems {\bf 85}, 55 (2006).

\bibitem{DOU04}
R.~J. Douglas and K.~A.~C. Martin: {\em Neuronal circuits of the neocortex\/},
  Annu.~Rev.~Neurosci. {\bf 27}, 419 (2004).

\bibitem{BOJ10}
I.~Bojak, T.~F. Oostendorp, A.~T. Reid, and R.~Kotter: {\em Connecting mean
  field models of neural activity to eeg and fmri data\/}, Brain Topography
  {\bf 23}, 139 (2010).

\bibitem{ZHO06}
X.~A. Zhou, G.~B. Qian, and S.~S. Qiu: {\em Chaotic control of nonlinear
  systems based on improving the space correlation\/}, Acta Physica Sinica {\bf
  55}, 3974 (2006).

\bibitem{PIN14}
D.~Pinotsis, P.~Robinson, P.~beim Graben, and K.~Friston: {\em Neural masses
  and fields: modeling the dynamics of brain activity\/}, Frontiers in
  Computational Neuroscience {\bf 8}, 149 (2014).

\bibitem{SAN15a}
P.~Sanz-Leon, S.~A. Knock, A.~Spiegler, and V.~K. Jirsa: {\em {Mathematical
  framework for large-scale brain network modeling in The Virtual Brain}\/},
  NeuroImage {\bf 111}, 385 (2015).

\end{thebibliography}
